\documentclass{eas}
\usepackage{graphicx}
\usepackage[authordate,bibencoding=auto,strict,backend=bibtex,natbib]{biblatex-chicago}
\usepackage{hyperref}
\usepackage[normalem]{ulem}
\usepackage{xcolor}
\usepackage{dirtytalk}
\usepackage{underscore}

\def\gtsima{$\; \buildrel > \over \sim \;$}
\def\ltsima{$\; \buildrel < \over \sim \;$}
\def\gtrsim{\lower.5ex\hbox{\gtsima}}
\def\lesssim{\lower.5ex\hbox{\ltsima}}

\begin{document}

%%-----------------------------
%%      the top matter
%%-----------------------------
\title{Stellar activity and transits} 
\author{G. Bruno}\address{INAF – Catania Astrophysical Observatory, Via Santa Sofia, 78, 95123, Catania, Italy}
\author{M. Deleuil}\address{Aix-Marseille Universit\'e, CNRS, CNES, LAM (Laboratoire d'Astrophysique de Marseille) UMR 7326, 13388, Marseille, France}%

\begin{abstract}
From an observational standpoint, stellar activity poses a critical challenge to exoplanet science, as it inhibits the detection of planets and the precise measurement of their parameters. Radial velocity and transit searches revealed a significant fraction of exoplanet hosts is active, and showed the need to fully understand the different facets of stellar activity and its impact on observables. Moreover, the activity correction is of prime importance for the detection and characterisation of Earth analogues.
We present a review of the effects that stellar activity features such as starspots, faculae, and stellar granulation have on photometric and low-resolution spectroscopic observations of exoplanets, and discuss the main aspects of the techniques which were developed to reduce their impact. 
\end{abstract}
\maketitle
%%-----------------------------
%%      your text
%%-----------------------------
\section{Introduction}
Our understanding of planetary systems is rooted in the knowledge of host stars' parameters and characteristics. Starting from the protoplanetary disc phase, stellar properties affect the types of planets that will form, as well as their composition \citep[e.g.][]{madhusudhan2017,mordasini2018}. As systems age, dynamical and magnetic interactions, together with stellar irradiation, are thought to importantly affect the evolution of planetary bodies \citep[e.g.][]{lanza2018}.

The precision obtained on stellar parameters also has a direct impact on the measure of planetary parameters, namely their masses and radii, which are derived relatively to the same parameters of their hosts \citep[e.g.][]{santos2018}. In turn, the masses and radii of exoplanets are necessary to calculate their mean densities, model their interiors, interpret their transmission spectra and compute atmospheric pressure scale heights \citep[e.g.][]{seager2007,guillot2014,dorn2015,brugger2017}.

While from one side stellar parameters must be determined with high accuracy and precision, stochastic stellar brightness variations also hamper the precision on measured quantities and prevent a correct determination of planetary parameters and properties. This might result in biases in planetary radii, masses and atmospheric composition.

In this review, the issue of stellar activity from the point of view of photometric and low-resolution transit observations is presented and discussed. In Section \ref{stellaract}, we introduce some of the main stellar activity indicators which are used for exoplanet surveys. In Section \ref{variations}, we show some benchmark cases of active stars observed in the context of exoplanet transit surveys, as well as the main techniques used to constrain the properties of starspots and faculae from observations. Sections \ref{phototransits}, \ref{transm}, and \ref{otherpar} are dedicated to the effect of stellar activity features on transits observed in photometry and low-resolution spectroscopy, and the impact of stellar granulation is the focus of Section \ref{granulation}. Section \ref{concl} presents some conclusions and perspectives which can be drawn from what we have learnt so far.

\section{Solar and stellar activity observations}\label{stellaract}
\begin{figure}
    \centering
    \includegraphics[scale=0.1]{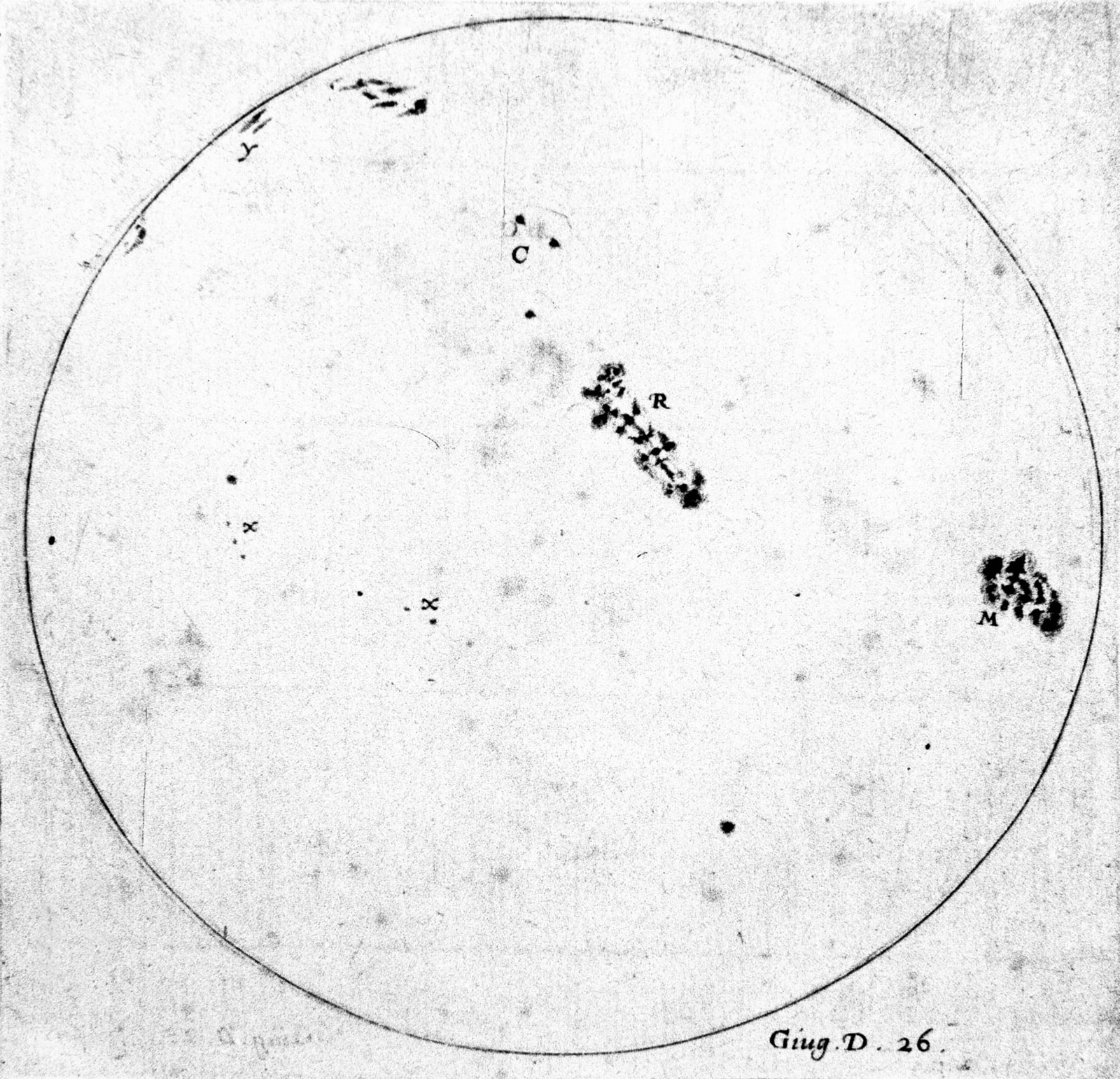}
    \caption{Illustration from Galileo's \textit{Istoria e dimostrazioni intorno alle macchie solari e loro accidenti} (``History and Demonstrations Concerning Sunspots and Their Properties'', or ``Letters on Sunspots''), 1613. Downloaded from Encyclop\ae dia Britannica \citep[][\url{https://www.britannica.com/science/sunspot}]{pallardy2012}.}
    \label{galileo}
\end{figure}

The first observations of stellar brightness variations come from the best-studied star we know, the Sun. Hints of the presence of darker regions on the solar disc, which we now call ``sunspots'', were recorded in China around 800 BC. In the western world, we can find the oldest records in the drawings in John of Worcester's \textit{Chronicles}, dating 1128. Thanks to the pioneering use of the telescope for astronomical observations, Galileo Galilei spotted inhomogeneities on the solar disc and recorded them on a plate dating 1612, as shown in Figure \ref{galileo}. The observation of stellar activity phenomena in the visible part of the spectrum, which were later extended to a broader range of wavelengths, enabled also the first detection of a solar flare by R. C. Carrington in 1859.

\begin{figure}
    \centering
    \includegraphics[scale=0.5]{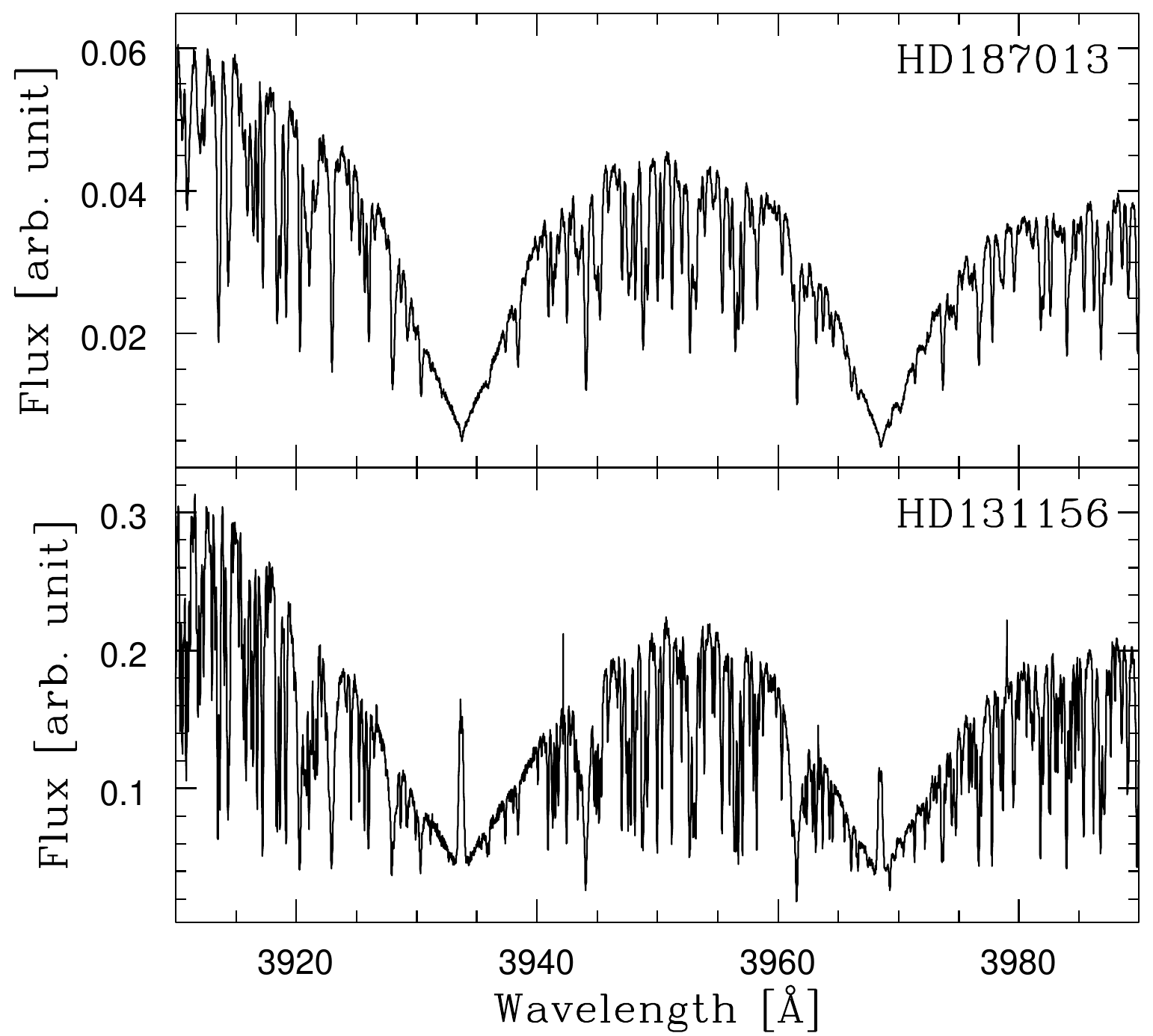}
    \caption{The CaII H and K lines in the SOPHIE spectrum of a non-active (\textit{top}) and active (\textit{bottom}) star. From \citet[][]{boisse2010}, A\&A, 523, A88, p. 9 (their Figure C.1), reproduced with permission \textcopyright ESO.}
    \label{calcium}
\end{figure}

Pioneering spectroscopic observations of CaII H\&K line emission at 393 and 397 nm in solar-type dwarfs were obtained by \citet{wilson1968,Noyes1984}. Time variations were shown for other spectroscopic lines \citep[e.g.][]{livingston2007}, such as H$\alpha$. Figure \ref{calcium} presents CaII H\&K lines of a non-active and an active solar-type star, whose spectral line core is affected by strong emission, seen as an indicator of stellar magnetic activity. These activity-related features revealed a dependence of the chromospheric activity level on the spectral type and the rotation period of the star \citep{wright2004}. Main-sequence cool dwarfs have their rotation braked due to the loss of angular momentum through a magnetised stellar wind. Hence, rotational velocity generally decreases with age, as well as activity \citep{gallet2013,salabert2016}, even if cases were detected of possible weakened magnetic braking and anomalously rapid rotation at old ages \citep{vansaders2016}. All in all, chromospheric activity is often used as an indicator of age \citep[e.g.][]{Pace2004,metcalfe2017}.

Photometry also provides indications of time-dependent, stellar brightness variations due to the passage of dark starspots and bright faculae across stellar discs. The amplitude of such variations clearly correlates with spectroscopic indicators \citep[e.g.][]{baliunas1995}. Additionally, observations provided evidence that the level of activity decreases with stellar age, corresponding to a transition from spot-dominated to facula-dominated stars \citep[e.g.][]{lockwood2007,preminger2011,shapiro2016,radick2018,reinhold2019}. In this regard, our Sun belongs to the category of weakly-active stars \citep{lockwood1997}. Photometric modulations are observed on various time scales, from days or tens of days -- indicating starspots and faculae brought about by stellar rotation -- to months or years -- revealing the evolution of the brightness dishomogeneities pattern on the stellar disc, activity cycles, and variations in the stellar disc coverage of activity features, or ``filling factor'' \citep[e.g.][]{nielsen2019}. This information enables placing constraints on stellar rotation rates, and on the distribution and evolution of activity features, and is therefore particularly helpful to understand the underlying dynamo mechanism across different spectral types \citep[e.g.][]{hall1991,lanza1998,berdyugina2004,berdyugina2005,strassmeier2005,strassmeier2009,bradshaw2014,see2016,brun2017,herbst2021}. 

In order to monitor daily photometric variations, ground-based observations are however not sufficient. Day-night alternation forces continuous interruptions, with adverse weather and difficulty to obtain continuous, long-term telescope time worsening the issue. In order to assess the impact of stellar brightness variations on transit observations, long-term uninterrupted monitoring is necessary. Hence, both exoplanet and stellar activity science greatly benefited from the launch of space telescopes dedicated to ultra-high precision and long-duration photometric observations, and better characterised the time scales concerned by stellar activity features \citep[e.g.][]{ballot2011,bradshaw2014}. Table \ref{spacemissions} gives a summary of these space-borne telescopes. 

\begin{table}[]
\small
    \centering
    \begin{tabular}{l|c|c|c|c|l}
        \hline \hline
        Telescope & Precision & Time sampling & mag limits & Sample size  & Reference\\
        & & & & [stars] & \\
        \hline
        \textit{SOHO/VIRGO} & $<10$ ppm/min &60 s & -- & The Sun & \citet{frohlich1995},\\
        & photon noise & & & & \citet{frohlich1997},\\
        & & & & & \citet{jimenez2002} \\
        \textit{MOST}    &  $\sim 100$ ppm/hr  &  $1-60$ s    &  $0.4 < V < 6 $     &    5000    &   \citet{walker2003}\\
        \textit{CoRoT} & 700 ppm/hr & 32 or 512 s & $11.5 < V < 16$ & $170000$ & \citet{auvergne2009}\\
        \textit{Kepler/K2} & 80 ppm/hr & 59 s or 29 min & $7 < K_{\rm p} < 17$ & $\sim 170000$ & \citet{koch2010}\\
        \textit{TESS} & 60 ppm/hr & 120 s or 30 min & $4 < V < 12$ & $> 200000$ & \citet{ricker2014}\\
        \textit{CHEOPS} & 50 ppm/hr & $1.05-60$ s & $6 < V < 13$ & $>600$  & \citet{broeg2013}\\
        \textit{PLATO} & 27 ppm/hr (goal) & 2.5 or 25 s & $8 < V < 16$ & $> 10^6$ & \citet{rauer2014}\\
        \textit{Ariel} & 10-100 ppm/hr (goal) & -- & bright stars & $\sim 1000$ & \citet{tinetti2018}\\
        \hline
    \end{tabular}
    \caption{Space missions dedicated to ultra-high precision stellar photometry and spectrophotometry.}
    \label{spacemissions}
\end{table}

\section{High precision photometry}\label{variations}
\begin{figure}
    \centering
    \includegraphics[scale=3]{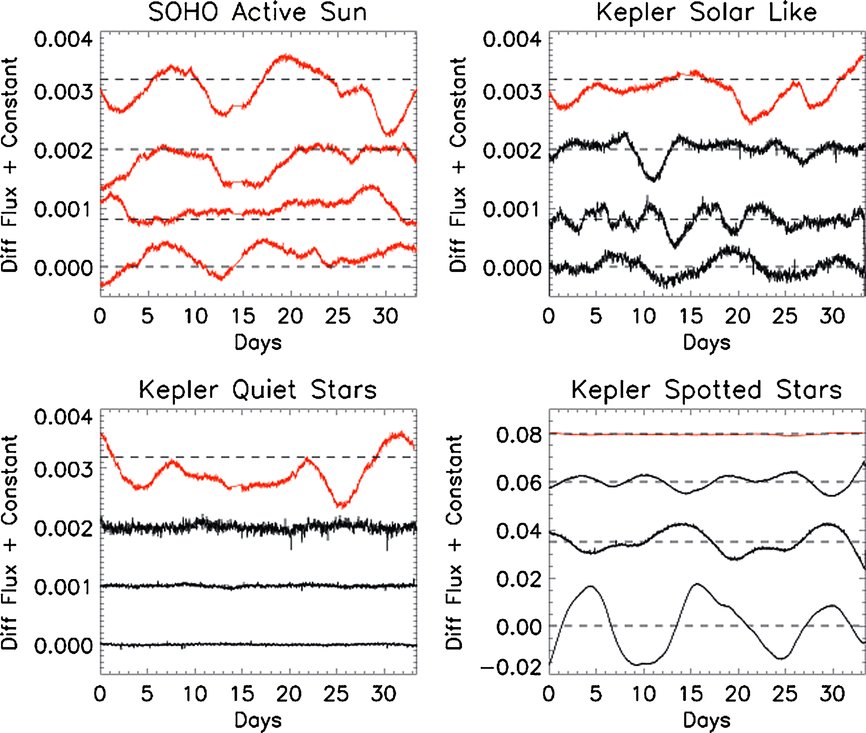}
    \caption{Examples of solar light curves (\textit{upper left}) compared with sample \textit{Kepler} light curves for solar-like (\textit{upper right}), quiet (\textit{lower left}), and spotted (\textit{lower right}) stars. The normalised light curves are offset with respect to each other for clarity. Solar light curves are coloured in red, and \textit{Kepler} light curves in black. In each panel, a solar light curve is shown for comparison. An offset was applied to each light curve. From \citet{basri2010}, ApJ, 769, 37, pp. L155-L159 (their Figure 2). Reproduced by permission of the AAS.}
    \label{basri}
\end{figure}

Space missions provided ultra-precise long-duration observations of hundreds of thousands of star, covering nearly all spectral types. However, even for the more than four year-long duration of the \textit{Kepler} observations, the duration of these missions is too short to cover the entirety of a stellar cycle (which, for the Sun, lasts about 11 years). Despite the limited temporal coverage, it was possible to observe a wide variety of brightness variation patterns and to connect them to stellar types. Figure \ref{basri} juxtaposes some light curves of the Sun to those of some \textit{Kepler} stars grouped by activity level. Comparative studies of this kind placed the Sun among weakly-active stars, or in a transition phase between active and non-active stars \citep{basri2010,basri2013,bohmvitense2007,brandenburg2017,strugarek2017}. Moreover, observations on as much as $\sim 10^5$ stars confirmed expectations that cool, mainly convective stars exhibit larger photometric variations \citep[e.g.][]{garcia2014}.

\begin{figure}
    \centering
    \includegraphics[scale=0.32]{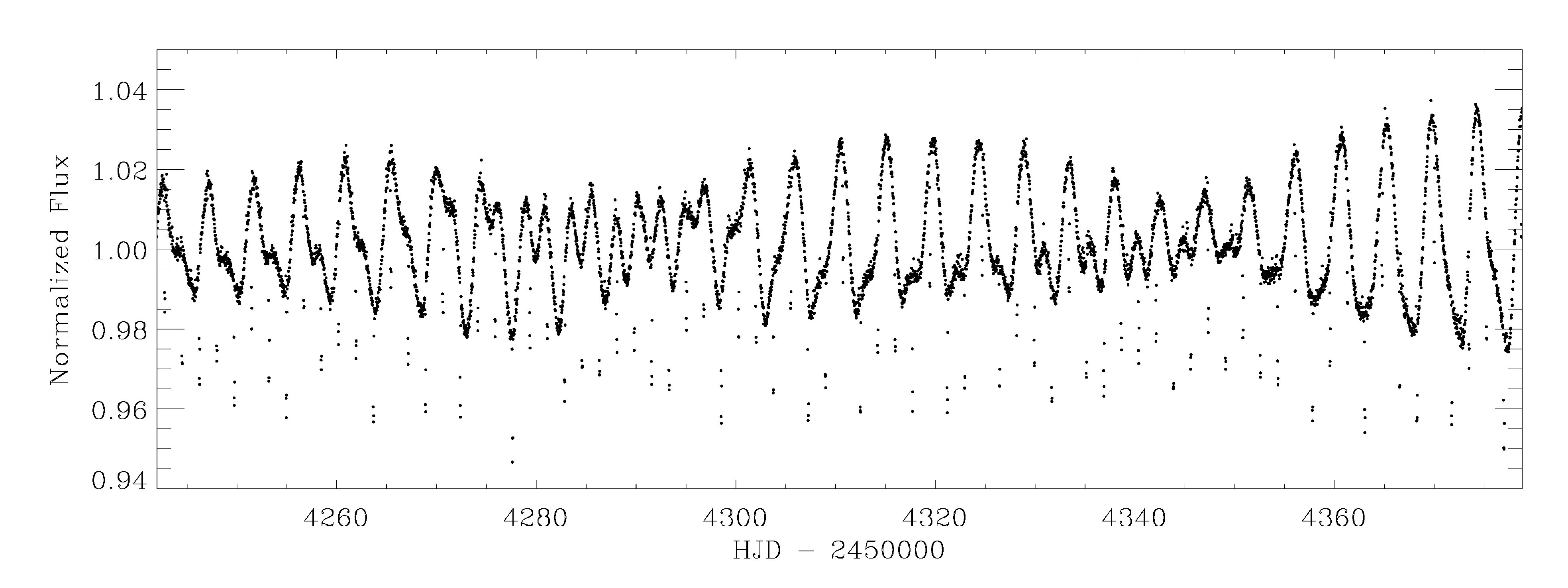}
    \caption{The light curve of CoRoT-2, where the transits of the planet are evident as vertical flux drops. In cases like this, the out-of-transit stellar flux cannot be determined with high confidence. Credit: \citet{alonso2008}, A\&A 482, L21-L24, p. 3 (their Figure 1), reproduced with permission \textcopyright ESO.}
    \label{alonso}
\end{figure}

This flux modulations observed in long-term and continuous light curves can reveal surface feature topologies and properties. As an example, Figure \ref{alonso} shows the light curve of CoRoT-2, hosting one of the first exoplanets discovered by the \textit{CoRoT} mission. From photometric amplitude variations, the size and brightness contrast (i.e., temperature) of starspots and faculae can be extracted. Deviations from a simple sinusoidal shape, which would be an indication of a single activity feature, increase as the number of stellar activity features increases and as their shape gets more complex. We have to remember, however, that we are not able to resolve the stellar disc, so that we cannot distinguish between large stellar activity features and agglomerations of smaller features. In addition, the complex inter-dependence of rotation, differential rotation, and the magnetic field-induced features that evolve in time, produces degenerate cases that hamper the inversion \citep[e.g.][]{mosser2009}. Despite this, long-term data series that allow a time coverage of a few stellar rotations have triggered new developments, bringing new insights into the configuration and evolution of starspot and faculae, but also key constraints on the stellar rotation period and differential rotation. 

\subsection{Stellar rotation periods}
Spots or inhomogeneities on the stellar surface are dragged away by the rotation of the star, resulting in a modulation of its brightness over days or tens of days. This modulation can be used to infer its rotation period, and was taken advantage of by photometric surveys which made large samples analyses possible \citep[e.g.][]{basri2010,affer2012}. 

By applying the autocorrelation function (proven to be more robust than the Lomb-Scargle periodogram for this kind of problem), \citet{mcquillan2014} were able to measure the rotation period of more than 34~000 stars, over a sample of more than 133~000 stars. Among other conclusions, they reported an increase of the rotation period with decreasing stellar mass and temperature, broadly consistent with the stellar spin-down law. Their findings, later expanded to an even larger sample \citep[e.g.][]{lanzafame2019,savanov2019,santos2019} also suggest that cooler stars are more evenly covered in brightness inhomogeneities such as starspots -- a factor that is particularly crucial to discern the effect of stellar activity on exoplanet transits. This aspect will be better explored in Section \ref{transm}.

\subsection{Stellar surface reconstruction} \label{inversion}
Knowledge on the distribution of stellar surface inhomogeneities can provide very valuable insights on the amount by which transit parameters are affected. Luckily, photometric light curves encode information on the distribution and temperature of such features. The complex problem of light curve inversion has been tackled in a number of ways, since much earlier than it became relevant to exoplanet science \citep[e.g.][]{budding1977,eaton1979,vogt1981,rodono1986}. Here we report some of the main approaches that have been used in the literature.

\begin{figure}
    \centering
    \includegraphics[scale=0.3]{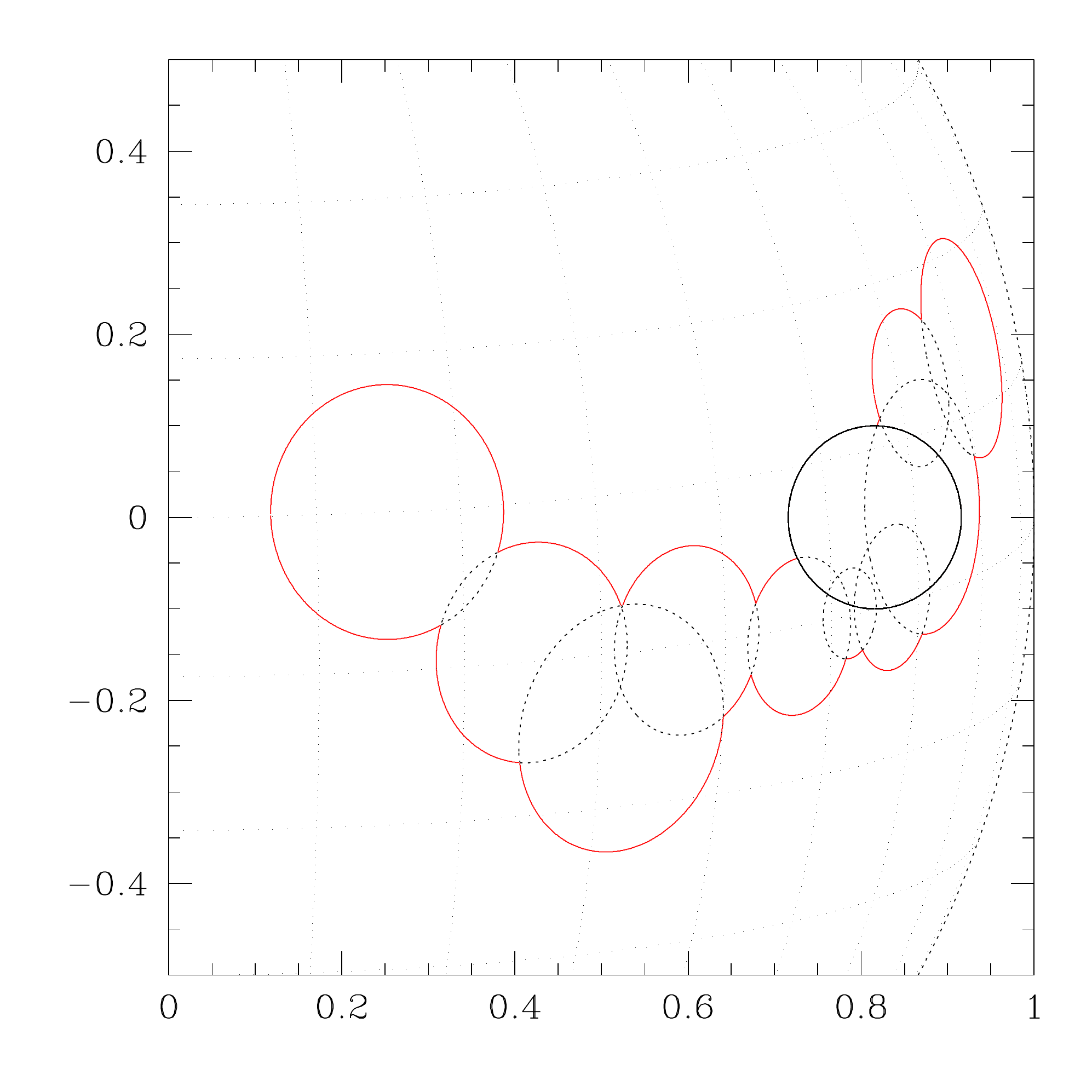}
    \includegraphics[scale=0.3]{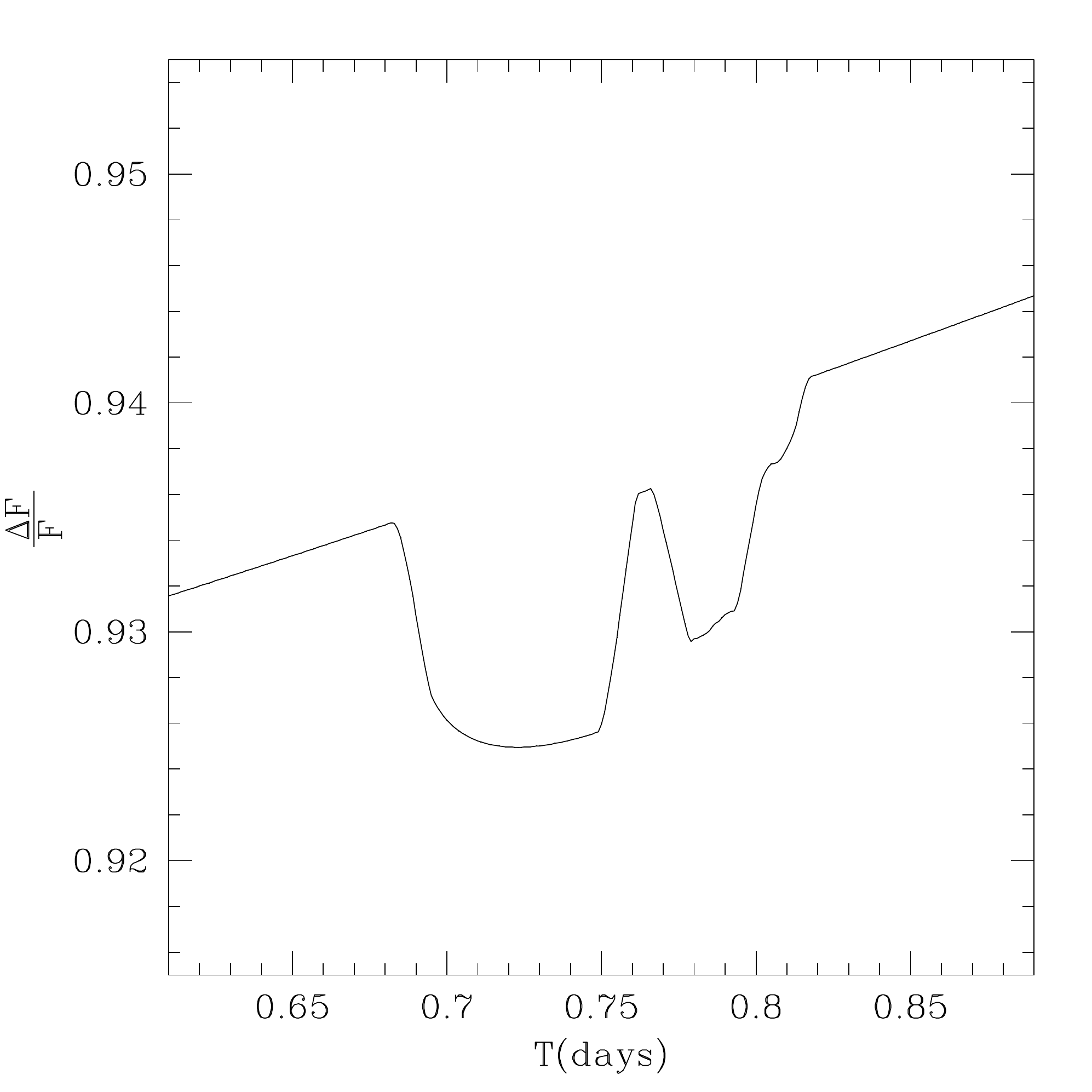}
    \caption{\textit{Left:} Geometric representation of a planet (black) transiting in front of a group of active regions (red). The axes denote the position on the star, normalised to the stellar radius, and are centred on the centre of the stellar disc. \textit{Right:} Corresponding normalised light curve. From \citet{montalto2014}, MNRAS 444, 1721–1728 (their Figure 2).}
    \label{montalto}
\end{figure}

\begin{figure}
    \centering
    \includegraphics[scale=1.6]{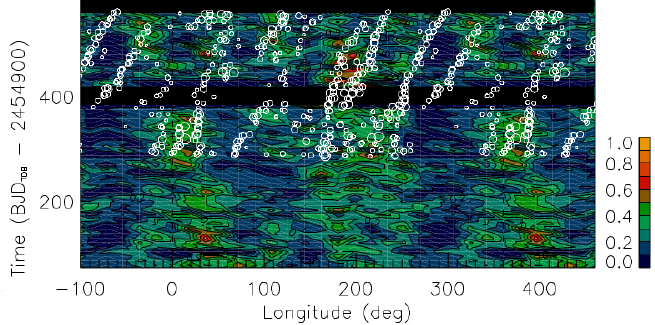}
    \caption{Reconstruction of the active region longitudinal distribution ($x$-axis) using hundreds of days of Kepler-17 photometric data (time on the $y$-axis). Modified from \citet{lanza2019}, A\&A 626, A38 (their Figure 9). The colours, from blue to orange, indicate an increasing filling factor, and identify active longitudes. Overplotted in white, the active region distribution identified by \citet{valio2017} using occulted starspots (see Section \ref{occulted}). Reproduced with permission \textcopyright ESO.}
    \label{lanza}
\end{figure}

\begin{itemize}
    \item Analytic models describe the stellar disc as the visible part of a sphere emitting a total unit flux, from which the contribution of darker (or brighter) regions with circular shape is subtracted (or added) to calculate the light curve \citep{budding1977,dorren1987,eker1994,kipping2012,montalto2014,beky2014,ikuta2020}. Usually, brightness inhomogeneities are distorted according to their projection at a given stellar longitude and latitude. Other parameters can be tuned, such as the inclination of the stellar rotation axis, the stellar rotation period, or the stellar limb darkening coefficients. Such a ``forward'' model is then implemented in a Markov Chain Monte Carlo (MCMC) algorithm (or its variations, such as Nested Sampling) to find the posterior probability distributions for the model parameters  \citep[e.g][]{kipping2012,bruno2016,ikuta2020}.\footnote{Among publicly available codes: \texttt{macula} (\url{https://github.com/davidkipping/MACULA}, \cite{kipping2012}), \texttt{KSint} (\url{http://eduscisoft.com/KSINT/}, \cite{montalto2014}), \texttt{Prism} (\url{https://github.com/JTregloanReed/PRISM_GEMC}, \cite{tregloan-reed2013}), and TOSC (\url{http://slnweb.oact.inaf.it/tosc/}, \cite{scandariato2017}).}\\
    An analytic model requires only a fraction of a second to be calculated and, in principle, there is no limit to the number of stellar activity features that can be modelled. Clearly, the more complex the model, the longer the computation time required to reach convergence. It is then often desirable to adopt some simplifications, as degenerate cases among parameters might hamper the convergence of the chains. In particular, not all models allow the modelling of planet-starspot occultations, which provide stong constraints on the active region distribution along the transit chord (see Section \ref{occulted}). Dedicated modelling approaches were developed for this specific problem \citep{tregloan-reed2013,beky2014,montalto2014,scandariato2017,juvan2018}. As an example, Figure \ref{montalto} represents the modelling of a planet occulting a particularly complex combination of bordering active regions.\\
    However, observations provide no constraint on some of the parameters. To overcome this issue, the temperature of the features can be fixed by adopting results from the literature \citep[e.g.][]{berdyugina2005}, or the stellar rotation axis can be fixed when there is no constraint on such parameter. Moreover, a large number of starspots or faculae might overcomplicate the model, and prevent a proper sampling of the parameter space by the MCMC.
    \item Alternatively, the stellar visible disc can be modelled with a grid of hundreds of elements, and the contribution of each cell to the total flux can then be evaluated. This allows the modelling of complex shapes for starspots and faculae, but is computationally very expensive \citep[e.g.][]{boisse2012,oshagh2013}.\footnote{In particular, the \texttt{SOAP 2.0} code \citep{boisse2012,oshagh2013} allows such a modelling, and is discussed in other proceedings of the EES2019. The code was presented by M. Oshagh at the school (\url{https://ees2019.sciencesconf.org/resource/page/id/9}) and can be downloaded at \url{http://www.astro.up.pt/resources/soap-t/} (installation instructions at \url{https://ees2019.sciencesconf.org/resource/page/id/8}).}\\
    The activity feature distribution can be retrieved in different ways. One is Maximum Entropy Regularisation, where  the brightness contrast (or the effective temperature) of starspots and faculae for each cell of the grid is fixed, and a functional that models the light curve given the filling factor $f$ of the cells (i.e., their fraction covered by activity features) is minimised:
    \begin{equation}
        \Theta = \chi^2(f) - \lambda S(f),
    \end{equation}
    where $\chi^2$ is the usual chi-square, $S$ is the entropy functional (a function describing the complexity of the model, so that more complex models are penalised) and $\lambda$ is a Lagrange multiplier. Despite the large number of free parameters, stable solutions can be achieved with this method, which retrieved active longitudes from some \textit{CoRoT} and \textit{Kepler} light curves of exoplanet-host stars \citep[e.g.][]{lanza2009,lanza2019}. Figure \ref{lanza} shows the reconstruction of the active region distribution using the light curve of Kepler-17 \citep{lanza2019}. Another option is to use a relatively small number of cells to describe the stellar disc: this allows one to adopt variations of the $\chi^2$ statistics, or even MCMC algorithms to achieve the best fit \citep[e.g.][]{wolter2009,huber2010,tregloan-reed2015}.
\end{itemize}

\subsection{Stellar differential rotation and starspot evolution}\label{diffrot}
In order to reconstruct the properties of the stellar dynamo and the generation of magnetic fields in the stellar convection zone, efforts have been directed to collect information on the variation of the stellar rotation period at different stellar latitudes \citep[e.g.][]{knobloch1982,brandenburg2005,balona2016}. This variation can be expressed as
\begin{equation}
    \Omega = \Omega_0 - \Delta \Omega \sin^2 \psi,
\end{equation}
where $\psi$ is the stellar latitude, $\Omega$ the rotation rate at the stellar equator and $\Delta \Omega$ the variation of the rotation rate between the stellar pole and the equator. The ``shear'' $\alpha = \Omega_0 / \Delta \Omega$ is one way to express the deviation of stellar rotation from the one of a rigid body, and is equal to 0.2 for the Sun ($\Delta \Omega_\odot = 0.055 \, \mathrm{rad \, day}^{-1}$).

\begin{figure}
    \centering
    \includegraphics[scale=0.55]{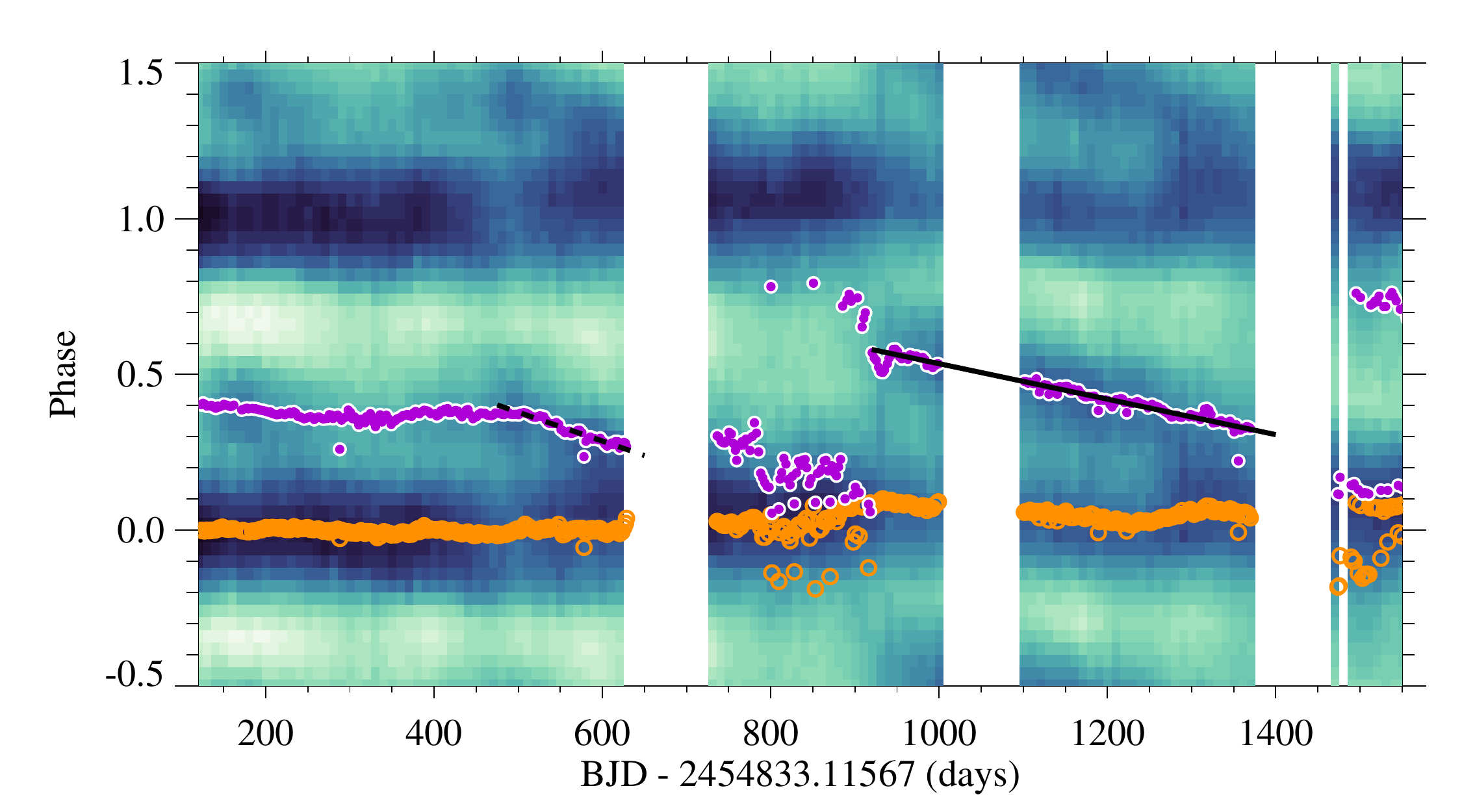}
    \caption{Reconstructed light curve map of GJ 1243, from \citet{davenport2015}, ApJ 806, no. 2, 212 (their Figure 5). The $x$-axis corresponds to the observation date, and the $y$-axis to the stellar rotation phase. The two retrieved starspots are marked in orange and purple. The orange pattern represents the higher-latitude, more stable starspot, and the purple one indicates the lower latitude feature migrating in longitude because of differential rotation. Reproduced by permission of the AAS.}
    \label{davenport}
\end{figure}

A variety of methods exist to estimate differential rotation from photometry and spectroscopy \citep[e.g.][]{berdyugina2005,lanza2014-differ,nielsen2015,santos2017}. The advent of space-borne high precision, long-term photometry added a powerful resource, both for the quantity of continuous data \citep[e.g.][]{lanza2014,garcia2014}, allowing to precisely track stellar rotation periods as a function of stellar latitude, and to asteroseismology \citep{benomar2018,bazot2019}. Here we discuss the example of \citet{davenport2015}, who applied a starspot modelling technique to reconstruct the longitudes and radii of starspots on the \textit{Kepler} light curve of the M dwarf GJ 1243. Using 5 days-long sliding windows and an MCMC sampler, they derived longitudes and radii of two starspot features: a high-latitude feature for over 6 years of observation, and a low-latitude feature evolving in both size and longitude over hundreds of days. For this second feature, they measured a differential rotation rate $\Delta \Omega = 0.012 \pm 0.002 \, \mathrm{rad \, day}^{-1}$ and a shear $\alpha = 0.00114$, one of the lowest values measured for cool stars. The phased light curve map is illustrated in Figure \ref{davenport}.

However, light curve modulations are degenerate with respect to the latitude of stellar activity features, so that these parameters were fixed during the modelling. As discussed in Section \ref{occulted}, the occultation of starspots and faculae during transits allows setting constraints on their latitudes, and hence provides significant information for better understanding the dynamo of the host star. Starspot and, for the first time, facula occultations were used to derive the differential rotation rate of a young solar-type star, Kepler-71, by \citet{zaleski2019}. The result of $\Delta \Omega < 0.005$ rad day$^{-1}$, much lower than that of the Sun's, describes a rigidly rotating star which is set apart from solar-type stars at a similar evolutionary state.

Another major limitation when trying to estimate the configuration of activity features is the description of their evolution, as most of the usual models are stationary. One way to address this issue is to split the light curve in chunks and to model them with different distributions of starspots and faculae \citep{silva-valio2010,lanza2011,davenport2015}. Indeed, stellar activity features do evolve in time, and characteristic laws can be used to describe their growth from first appearance to maximum size, permanence at maximum size and decay until disappearance \citep[e.g.][]{kipping2012,bradshaw2014,vandrielgesztelyi2015,savanov2019-spots,namekata2019}. With a starspot and facula model including feature evolution, longer segments of the light curve can be modelled and the lifetime of individual features (or groups) can be inferred \citep[e.g.][]{mosser2009,bruno2016}.

\section{Planetary transits and stellar activity features in photometry}\label{phototransits}
Stellar activity directly affects planetary transits in different ways, as spots and faculae leave their imprint on both their depth and shape. This impacts the observables used to derive the radius of an exoplanet, and as such it is of crucial importance to disentangle them from any kind of contamination.

\subsection{Effect of non-occulted activity features on the transit depth}\label{nonoccphot}
Variations in the apparent transit depth are caused by changes in the out-of-transit stellar brightness. If $F_\mathrm{out}$ is the out-of-transit stellar flux and $F_\mathrm{in}$ the in-transit stellar flux (both at the centre of the transit), the transit depth is defined as 
\begin{equation}
    D \equiv \frac{F_\mathrm{out} - F_\mathrm{in}}{F_\mathrm{out}} = \left( \frac{r_\mathrm{p}}{R_\star} \right) ^2,
\end{equation}
where $r_\mathrm{p}$ and $R_\star$ are the planetary and the stellar radius, respectively. If the star  is covered by activity features,  $F_\mathrm{out} - F_\mathrm{in}$ remains the same, but $F_\mathrm{out}$ changes with time: in particular, if it decreases (increases) because of starspots (faculae), the transit depth will increase (decrease). Brightness variations are partly accounted for by fitting the out-of-transit flux in proximity of the transit edges with a low-order polynomial, and by computing its value along the transit. The relative effect is more remarkable the larger the brightness modulations are, and the smaller the planet is. Variations can span from a few $10^2$ ppm to a few percent \citep{czesla2009}, with considerable consequences on the inferred planetary properties (such as its internal structure and composition, e.g. \cite{wagner2011}). 
\citet{czesla2009} suggested that a different quantity than the transit depth should be adopted in order to minimise the effect of stellar brightness variations. In their notation, $f \equiv F_\mathrm{out} - F_\mathrm{in}$, $n \equiv F_\mathrm{out}$, and the proposed metric is
\begin{equation}
   D = \frac{f - n}{p} + 1,
\end{equation}
where $p$ is the value of the stellar photometric flux unaffected by starspots or faculae. The problem with this expression, however, is that the contribution of faculae to the stellar flux is not known \textit{a priori}, so that the value of $p$ cannot be deduced from observations nor theory.

Another way to estimate the out-of-transit stellar flux is via the use of starspot and faculae modelling. If observations covering a few stellar rotations are available, the distribution of starspots and faculae, as well as their brightness contrast (i.e., effective temperature) can be reconstructed up to some level of confidence, and $F_\mathrm{out}$ at any intermediate time can be inferred. The precision on this value depends on the uncertainty on the retrieved activity feature distribution and its degenerate solutions. One can remove the transits from the light curve, describe the out-of-transit flux with one of the models outlined in Section \ref{inversion}, and use this value for the in-transit flux at any given time. Alternatively, models enabling simultaneous simulations of both activity features and transits can be fitted to the entire light curve. In this way, constraints on both the transit and the activity feature parameters can be obtained, and the need for transit normalisation avoided \citep{bruno2016}. The price for this is a considerable increase in computational and model selection effort for the transit fit. 

A third option, which has recently gained popularity, is to apply a Gaussian process (GP) regression to the whole light curve altogether, setting the transit as the mean function of the process \citep{aigrain2016,foreman-mackey2017,barros2020}.\footnote{A presentation dedicated to GP regression was given by S. Aigrain at the EES2019 (\url{https://ees2019.sciencesconf.org/resource/page/id/9)}.} With a GP, the stellar signal is modelled as correlated noise, without the need to describe it explicitly. However, this requires the choice of the functional form for the correlation among observations, i.e. a ``kernel''. Some specific kernels are often adopted to model the variations induced by starspots and faculae, such as combinations of the exponential-squared and exponential-sine-squared kernels \citep{haywood2014}. The GP can be calibrated so that the kernel parameters (or ``hyperparameters'') are used to perform posterior inference on the size, position, and contrast of active regions, as well as their evolution time scales \citep{luger2021-1,luger2021-2}. Convenient open-source software for the GP implementations here described is available online\footnote{For example, \texttt{George} \citep[\url{https://github.com/dfm/george}][]{ambikasaran2015}), \texttt{celerite} (\url{https://github.com/dfm/celerite}, \cite{foreman-mackey2017}) and the GP regressors in \texttt{scikit-learn} (\url{https://scikit-learn.org}, \cite{pedregosa2011}).} and can be readily included into an MCMC simulation. GPs are particularly powerful thanks to their flexibility, but they are also very demanding in terms of computation resources: given $N$ data points, this is $\mathcal{O}(N^3)$ for most kernels, and $\mathcal{O}(N)$ for a specific kind of kernels applied to one-dimensional data sets \citep[e.g.][]{ambikasaran2015,foreman-mackey2017}.

Which approach is the most advantageous depends on the case at hand. The simplest model might be the most useful when only a few transits are available (even only one, as it is sometimes the case). Analytic modelling of starspots and faculae is recommended to obtain a physical representation of the starspot configuration and temperature, given an uninterrupted light curve, and it can provide with a more robust assessment of the out-of-transit flux. GPs, on the other hand, avoid the need of precise parametrisation of the geometry of the problem and can be coupled with models of instrumental systematics with a relatively low number of free parameters. None of these methods is completely exhaustive, and the next Section focuses on how occultations of activity features can help break some parameter degeneracy. 

\subsection{Effect of occulted activity features}\label{occulted}
\begin{figure}
    \centering
    \includegraphics[scale=0.4]{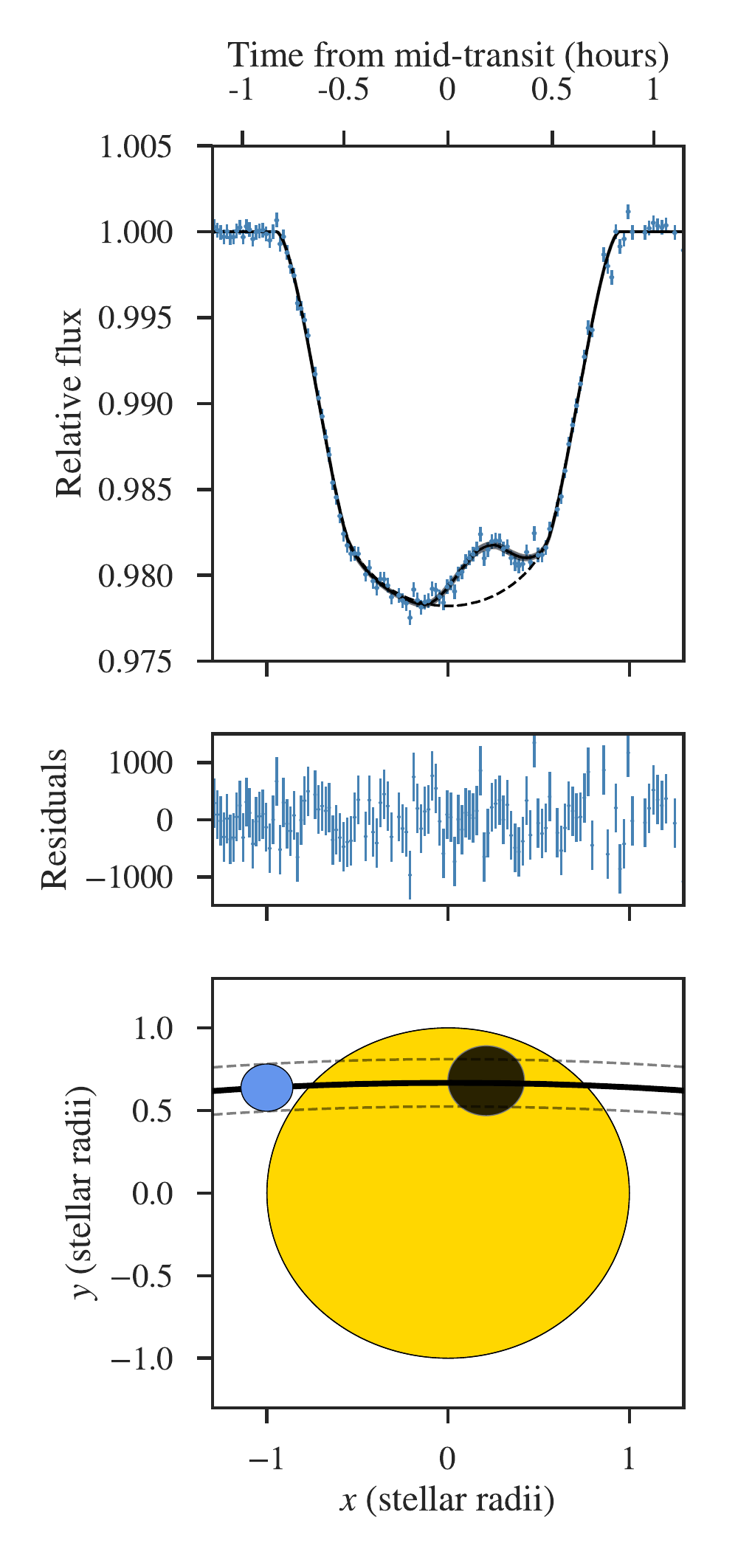}
    \includegraphics[scale=0.4]{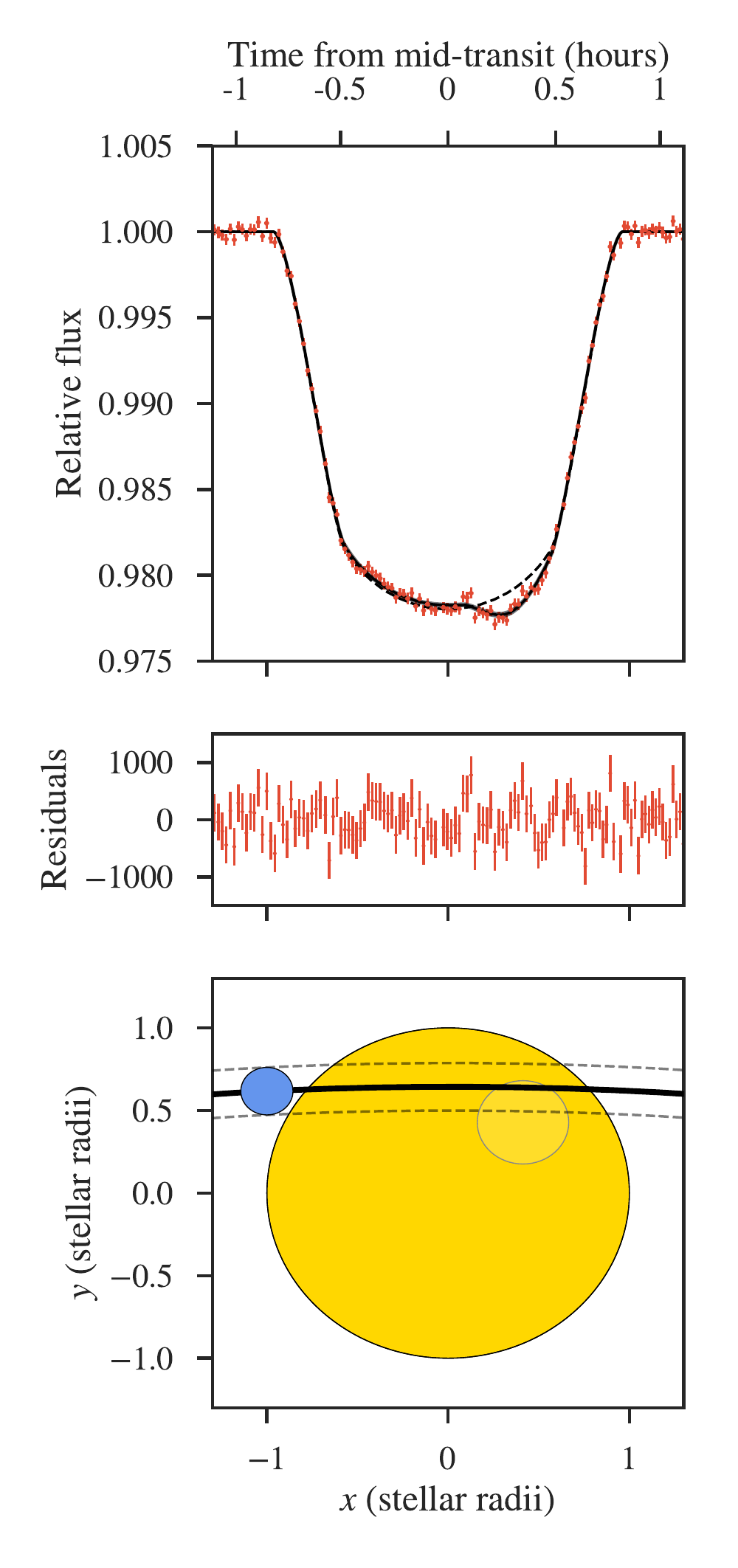}
    \caption{Effect of the planet crossing of a spot (\textit{left}) and a facula (\textit{right}) on the transit profile. From \citet{espinoza2019}, MNRAS 482, no. 2 (their Figure 7).}
    \label{espinoza}
\end{figure}

When starspots or faculae are occulted by a planet during a transit, they produce typical ``bumps'' on the transit profile. They affect transit depth measurements in an opposite way to non-occulted ones: if the planet passes in front of a dark starspot, the apparent transit depth will temporarily decrease, and the opposite will happen in case a bright facula is occulted (Figure \ref{espinoza}). From the Sun, we know that faculae are often associated to sunspots \citep{kiepenheuer1953}. They are mostly visible when they are close to the stellar limb, and are therefore described by a wavelength-dependent limb brightening law \citep[e.g.][]{spruit1976,shapiro2019}.

If the temperature contrast of the activity features with the average stellar photosphere is large enough, given the noise level of the data, their occultations can be easily identified in the transit profile. In principle, they can therefore be removed from the transit profile, which can then be fitted with a standard transit model. However, if their signal cannot be clearly singled out, or during a given single transit the planet crosses multiple stellar activity features, removing the imprint of the occultation might be problematic. The case of undetected occulted features was discussed by e.g. \citet{ballerini2012}, while the problems of multiple starspot crossings in the same transit were explored by e.g. \citet{czesla2009} and \citet{silva-valio2010} on the benchmark case of the hot Jupiter CoRoT-2b. These latter authors found that, under the assumption that all crossed features are dark starspots, the transit radius of CoRoT-2b might be overestimated by up to 3\% if the crossing effect is not taken into account. However, our knowledge of starspots and faculae in other stars is still not complete enough to exclude the contribution of faculae, which could average out this effect, or even increase the apparent transit depth in individual transits. Indeed, \citet{bruno2016} found indications that some of the deepest transits of CoRoT-2b, previously considered as the closest ones to the ``unperturbed'' transit profile, might be affected by faculae.

A very interesting advantage of starspot and facula crossings is that their position on the stellar disc (assuming the stellar rotation axis is known) can be determined with a much larger precision than in the case of non-occulted features. Provided a stellar inclination angle of $90^\circ$ with respect to the line of sight, the feature latitude $\phi$ can in fact be constrained within the stellar latitudes crossed by the transit chord:
\begin{equation}
\phi = \arctan \left( \frac{r_\mathrm{p}}{R_\star} \right).    
\end{equation}
Thanks to this, the latitudinal distribution of the starspots on HAT-P-11 was shown to follow a solar-like behaviour \citep{morris2017}.

Occultations also provide precise constraints on the longitude of active regions, which are found to be consistent with the reconstruction of the stellar surface using only unocculted features \citep[][see Figure \ref{lanza}]{lanza2019}. In the special case of Kepler-7b, \citet{desert2011} were able to observe the recurrent passage of the same group of occulted starspots on the transit profile. Together with a previous measurement of the stellar obliquity via the Rossiter-McLaughlin effect, and helped by the commensurability of the stellar rotation and planet revolution period ($P_\star/P_\mathrm{planet} \simeq 8$), it was possible to follow the starspots moving across different transits. Thanks to the long duration of the Kepler-7 light curve, starspot groups were observed without interruption for about 100 days, and important constraints on the lifetime of these features were obtained.

A last aspect to underline is the possibility of constraining the size and temperature of spots and faculae thanks to their occultation. A minimum value on their size can be inferred from the duration of the bump they produce on the transit profile, and this can be scaled to a feature size if the transit duration and stellar radius are known. With a simple geometric model that takes the activity feature brightness contrast with the stellar surface as a free parameter, one can derive the temperature contrast of the active feature to the stellar photosphere by inverting the ratio between black body laws \citep{silva2003}:
\begin{equation}
    \rho = \frac{\exp{(h \nu /K_\mathrm{B} T_\mathrm{eff})} - 1}{\exp{(h \nu /K_\mathrm{B} T_\mathrm{feat})} - 1},
\end{equation}
where $\rho$ is the brightness contrast with the stellar photosphere, $h$ is Planck's constant, $K_\mathrm{B}$ is Boltzmann's constant, $\nu$ is the frequency of observation and $T_\mathrm{eff}$ and $T_\mathrm{feat}$ are the stellar and the activity feature effective temperature, respectively. This method was adopted in several cases \citep[e.g.][]{silva-valio2010,mancini2017}.

Clearly, the black body law is only a useful approximation as stars, starspots, and faculae are not black bodies at different temperatures, and each one has its own spectrum and specific absorption lines. Section \ref{transm} contains a discussion on how transmission spectroscopy can help us probe not only planetary spectra, but also those of occulted stellar activity features.

\section{Starspots and faculae in transmission spectroscopy}\label{transm}
Transiting exoplanets are key targets to probe the physical structures and chemical compositions of their atmospheres. As their upper atmosphere absorbs and scatters starlight, transmission spectroscopy can be used to infer the physical conditions within their atmospheres by observing the transit depth at different wavelengths \citep{charbonneau2002}. Here, also, the impact of starspots and faculae cannot be neglected because, as discussed in Section \ref{nonoccphot}, they increase and decrease the measured transit depth, respectively. 
The variations are not only time-dependent, but also wavelength-dependent: according to Wien's law, the peak emission from a black body moves to longer wavelengths as the black body gets cooler, and there is more sensitivity close to the emission peak. Hence, the effect of starspots and faculae is stronger in the visible, and less prominent (but still important) in the infrared (IR). 

\begin{figure}
    \centering
    \includegraphics[scale=10.0]{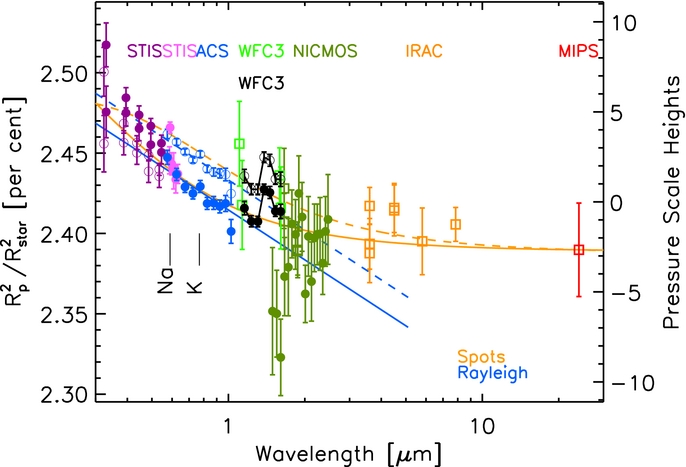}
    \caption{The transmission spectrum of HD 189733 b in different bands. In the visible part of the spectrum, blue and orange colours represent Rayleigh scattering and unocculted starspot models, respectively. From \citet{mccullough2014}, ApJ 791, 55 (their Figure 4). \textcopyright AAS. Reproduced with permission.} 
    \label{mccullough}
\end{figure}

\begin{figure}
    \centering
    \includegraphics[scale=0.7]{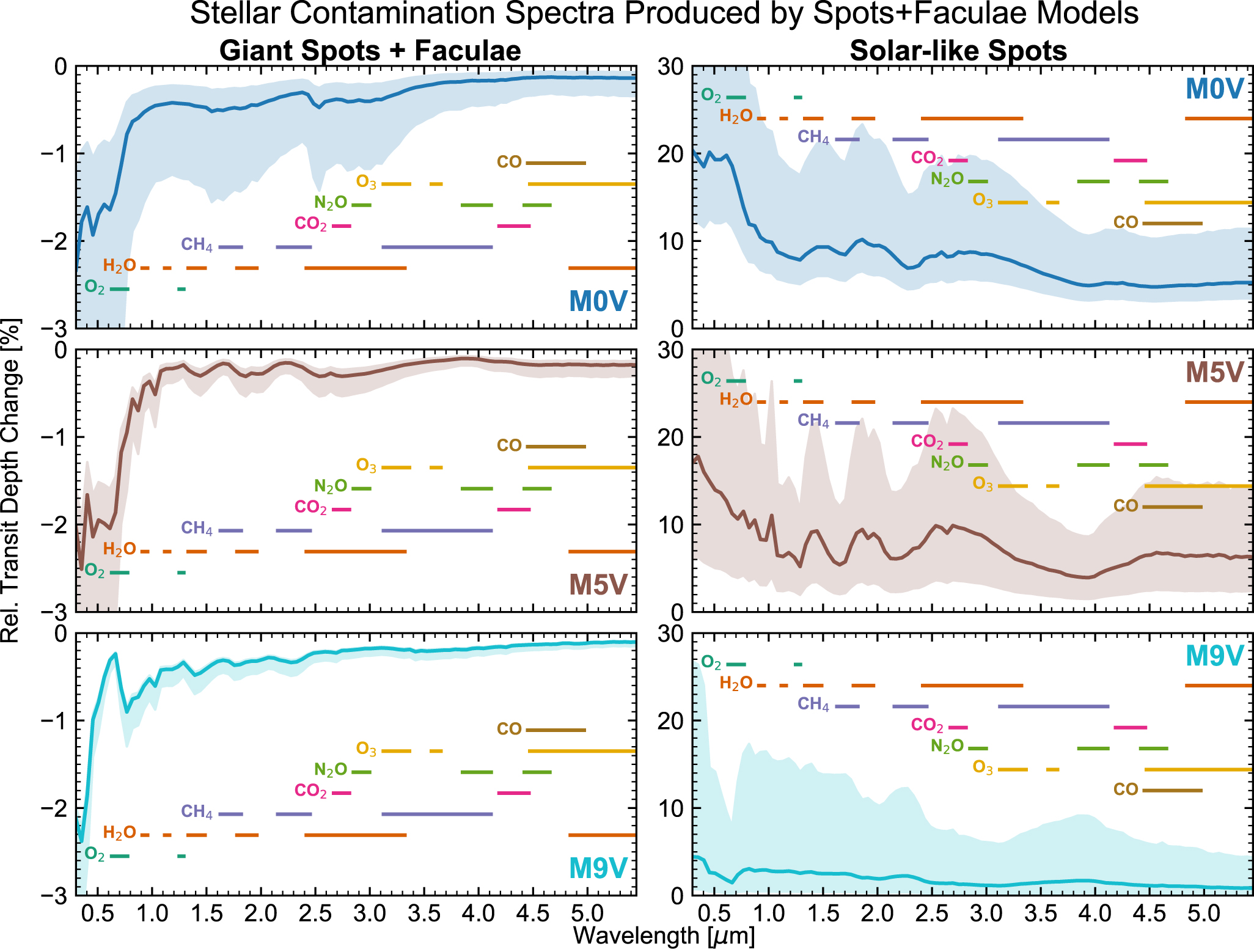}
    \caption{The amount of contamination of transmission spectra for planets hosted by active M stars, in the case of giant (\textit{left}) and solar-size starspots (\textit{right}), according to \citet{rackham2018}, ApJ 853, 122 (their Figure 7). The molecular absorption bands at various wavelengths are reported. Reproduced by permission of the AAS.}
    \label{rackham}
\end{figure}

As shown in the example of Figure \ref{mccullough}, it was found that unocculted starspots (faculae) result in a rising (descending) slope in the transit depth from the near-IR to the ultraviolet, which could be wrongly interpreted as Rayleigh-scattering dust (or its absence) in the planet’s upper atmosphere \citep[e.g.][]{pont2008,pont2013,mccullough2014,rackham2017,alam2018,sotzen2020}. The opposite happens for occulted active regions \citep{oshagh2014}. 

Among molecular spectral signatures which are nearly routinely detected, the water absorption band is the main one observed by \textit{Hubble Space Telescope}'s Wide Field Camera 3 instrument, and is the main indicator of oxygen abundance and metallicity in an exoplanet atmosphere. However, in starspots cooler than $\sim 3000$ K, water vapor could exist \citep{wallace1995} and produce absorption features that can mimic planetary water absorption at $\sim 1.4$ and $\sim 2.3 \, \mu$m \citep[e.g.][]{wakeford2019}. 

Because of this, an assessment of the activity level of the exoplanet host is necessary before any interpretation of a transmission spectrum is attempted. The correction of the activity-induced transit depth variations can be done by a photometric monitoring of the host star, with stellar models (such as \cite{castelli-kurucz2004} 1D ATLAS and PHONENIX \cite{husser2013}) to account for the wavelength-dependent flux correction, in order to properly evaluate the stellar contribution \citep[e.g.][]{huitson2013,alam2018}. 
In this kind of modelling, the size of the activity features was shown to be essential \citep{rackham2018}. With the use of stellar synthetic spectra to model both stars with different spectral types and their starspots, \citet{rackham2018} found that active regions with angular size comparable to the size of large sunspot groups ($\sim 2^\circ$, \cite{mandal2017}) can produce up to tens of percent variations in atomic and molecular absorption bands, against a few percent in the case of features which are four times as large. This is due to the fact that, for a given activity-induced stellar photometric variability amplitude, small active regions require much larger covering fractions to produce such amplitude, and produce a larger contamination in the transmission spectrum. Figure \ref{rackham} presents their findings on spectral contamination from starspots on M dwarfs. With the use of similar simulations, the same effects were shown to be much smaller for FGK stars \citep{rackham2019}. 

The shape of the out-of-transit light curve could be the key to distinguish giant from solar-like activity features on planet hosts, and so to identify optimal targets for transmission spectroscopy of planets orbiting cool stars. In the case of giant starspots, wide variations of the light curve will be observed, with a periodicity which depends on the stellar rotation period. In the case of small starspots, especially if these are uniformly distributed on the stellar disc, brightness variations will have a much smaller amplitude and no clear periodicity. This might be the most frequent case for M stars (see Section \ref{variations}) where, on average, a small starspot going out of view could be compensated by another one coming into view. The two scenarios could be discerned thanks to a continuous monitoring of the host star in the time windows during which transits are observed. Inspection of the light curve for at least two or three rotation periods in different spectral bands should enable to distinguish the type of light curve variations, and so to gauge the amount of contamination on the transmission spectrum. Remarkably, a time-evolving, longitudinal map and effective temperature reconstruction of the active regions on WASP-52 was obtained from $\sim600$ days of $BVRI$ multiband photometry, implying a reduction of their effect on transmission spectroscopy by about one order of magnitude \citep{rosich2020}. A complementary approach would be to model the temperature and size of active regions using out-of-transit low-resolution spectroscopic observations. This technique showed promising results for the correction of transmission spectra affected by large-contrast spots, in modelled bright ($V=9$) stars \citep{cracchiolo2021}.

The time dependence of the ``filling factor'' (i.e., the fraction of the stellar surface covered by activity features) also operates a variation in the baseline between two or more transit observations. The photometric monitoring of the host star's activity level was, in fact, shown to be key to correctly stitch together multi-epoch transmission spectroscopy observations, aiming either at extending the wavelength coverage or at detecting temporal variations of the planet’s atmosphere. Retrieval exercises including the contribution of activity features to the spectrum were also shown to provide constraints on both the atmospheric and the starspot parameters \citep[e.g.][]{barstow2015,rackham2017,zellem2017,bruno2020}. 

\begin{figure}
    \centering
    \includegraphics[scale=0.4]{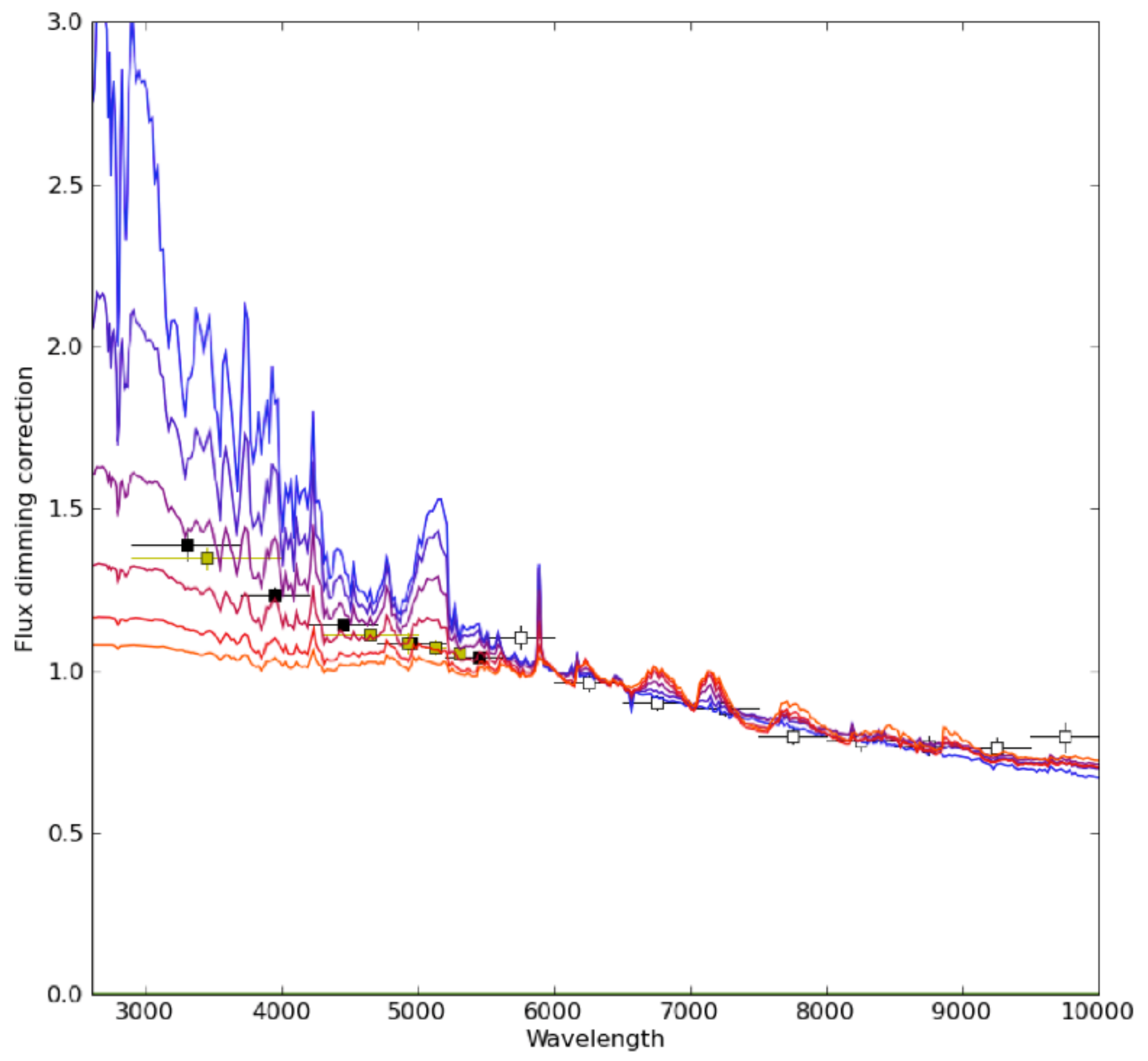}
    \caption{Relative height of the spot occultation bump with respect to the same quantity at 600 nm, as a function of wavelength (data points), and Kurucz 1D ATLAS models used to fit the measured flux rise during a transit of HD 189733b. Measurements (symbols) were obtained from \textit{Hubble Space Telescope}'s STIS G430L and ACS instruments. Starspot models range from 4750 to 3500~K in 250~K intervals (blue to orange, respectively), while a $T_\mathrm{eff} = 5000$~K ATLAS stellar model \citep{castelli-kurucz2004} was used for the host star. From \citet{sing2011}, MNRAS 416, 1443–1455 (their Figure 10).}
    \label{sing}
\end{figure}

Simultaneous transit observations in multiple wavelengths can provide very useful information on the physical properties of activity features when they are occulted by the planet. In that case, indeed, the relative flux increase (the height of the starspot occultation bump) $\Delta f$ measured in different wavelengths $\lambda$ probes specific absorption lines. This makes it possible to place constraints on the effective temperature of the activity feature, whose emission is modelled using stellar synthetic spectra. Following \citet{sing2011},
\begin{equation}
    \frac{\Delta f(\lambda)}{\Delta f(\lambda_0)} = \frac{1 - F_\lambda^{T_\mathrm{feat}}/F_\lambda^{T_\mathrm{eff}}}{1 - F_{\lambda_0}^{T_\mathrm{feat}}/F_{\lambda_0}^{T_\mathrm{eff}}},
\end{equation}
where $F_x^y$ is the normalised flux measured at wavelength $x$ and effective temperature $y$, and $\lambda_0$ is a reference wavelength. Figure \ref{sing} shows the so-determined spectrum of a starspot occulted during a transit of HD 189733b. For the specific case of this active K0-dwarf star, \citet{sing2011} derived a spot temperature $T_\mathrm{spot} \simeq 4250 \pm 250$~K  and spot-to-star temperature differences compatible with what is measured for sunspots. In the case of WASP-19, the fit of the occultation distortion at varying wavelengths allowed the determination of better than 100 K constraints on a bright and a dark active region \citep{espinoza2019}.

\section{Effects on other transit parameters}\label{otherpar}

\begin{figure}
    \centering
    \includegraphics[scale=0.3]{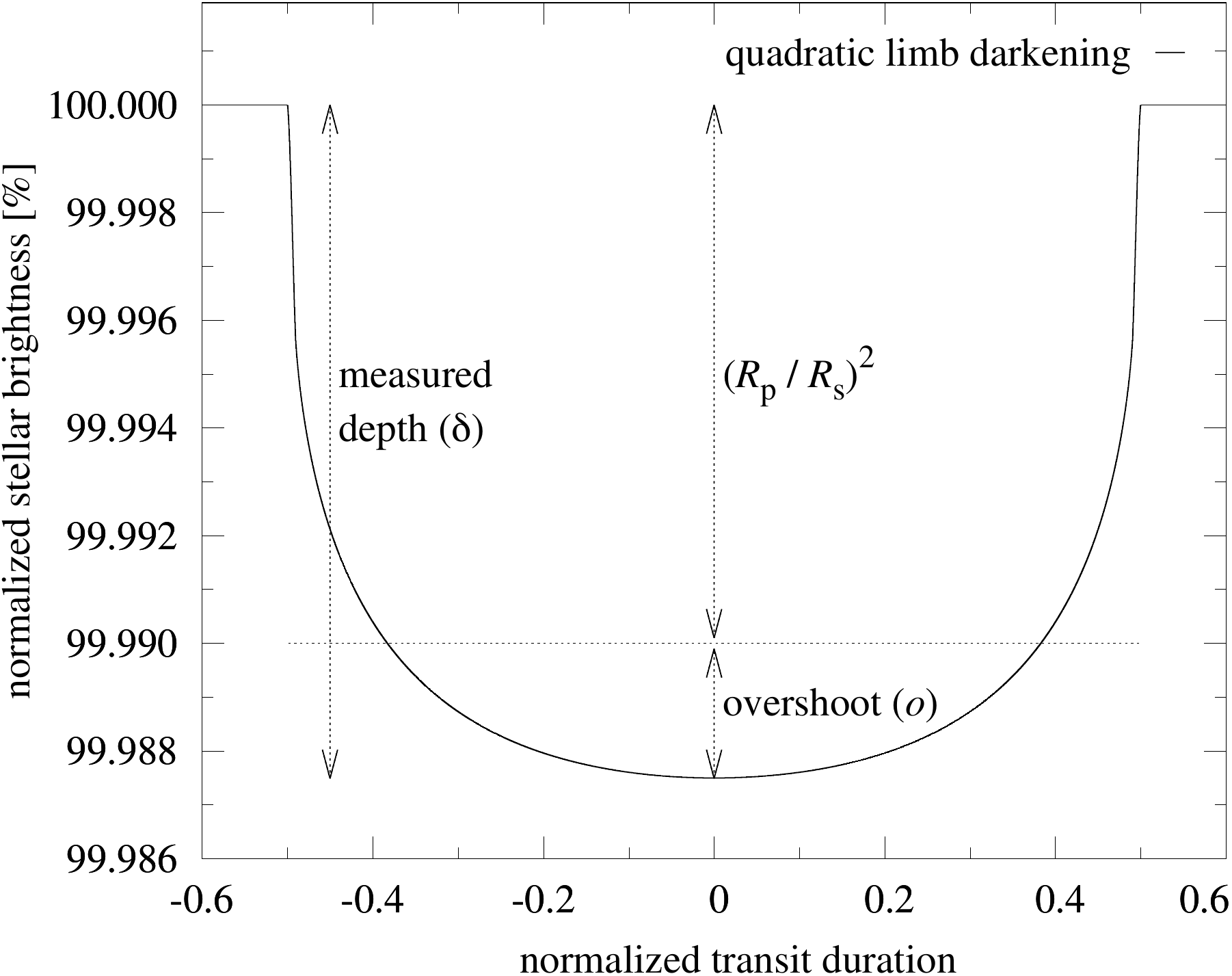}
    \includegraphics[scale=0.3]{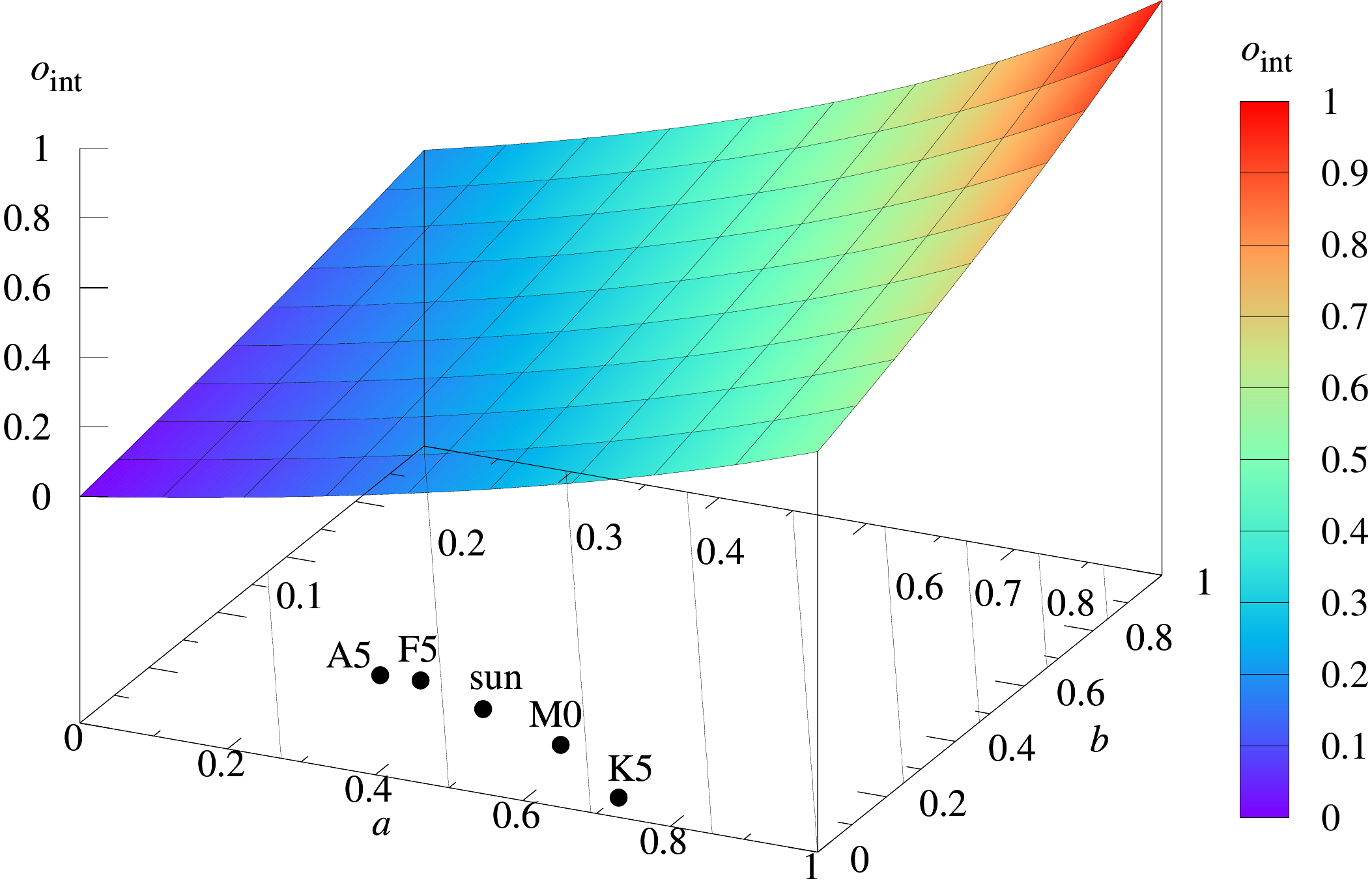}
    \caption{\textit{Left}: the effect of stellar limb darkening on the transit shape and depth. The increase in transit depth over the planet-to-star surface ratio is here called ``overshoot'' ($o$). \textit{Right:} The variation in overshoot ($o_\mathrm{int}$, $z$-axis) for a quadratic limb darkening law (coeffcients $a$ and $b$) and orbit inclination $i = \pi/2$ rad for different stellar types. Variations in the limb darkening coefficients due to stellar activity cause variations in the overshoot. From \citet{heller2019} (Figures 1 and 3), A\&A 623, A137, pp. 2 and 5, reproduced with permission \textcopyright ESO.}
    \label{heller}
\end{figure}

Among the transit parameters which are affected by stellar activity features, limb darkening (LD) is one of the most crucial. In turn, centre-to-limb stellar brightness variations  modify both the transit profile and its apparent depth, affecting the determination of the exact radius of the planet \citep{mandelagol2002,csizmadia2013,heller2019}. In Figure \ref{heller}, left panel, the transit profile shape variation is shown when the effect of LD is included. In the figure, the transit depth increase is dubbed ``overshoot''. On the right, the overshoot is analytically derived and shown for different spectral types. If stellar activity features affect the LD coefficients, the transit depth measure (and therefore the measured radius of the planet) for a given stellar type is prone to error.

Usually, LD coefficients are obtained by fitting stellar centre-to-limb specific intensity model profiles with a variety of increasingly complex analytic relationships. For example, two popular laws are the quadratic \citep{kopal1950} and the non-linear \citep{claret2000} LD law. Using $\theta$ as the angle between the stellar surface normal and the line of sight, they are parametrised as
\begin{equation}
    \frac{I_\lambda(\mu)}{I_\lambda(1)} = 1 - a_1(1 - \mu) - a_2(1 - \mu)^2,
\end{equation}
and
\begin{equation}
    \frac{I_\lambda(\mu)}{I_\lambda(1)} = 1 - \sum_{n=1}^{4} a_n(1 - \mu^{n/2}),
\end{equation}
respectively, where $\mu=\cos \theta$ and $I_\lambda(\mu)/I_\lambda(1)$ expresses the ratio between the specific intensity at any wavelength $\lambda$ and position $\mu$ and the centre of the stellar disc (corresponding to $\mu = 1$).  

The LD coefficients $a_n$ can be fitted to the stellar models and then used as fixed parameters to model the transit profile \citep{claret2000}. In particular, the non-linear law has been shown to produce the most accurate fit to the models so far \citep[e.g.][and references therein]{morello2017}. 

\begin{figure}
    \centering
    \includegraphics[scale=0.34]{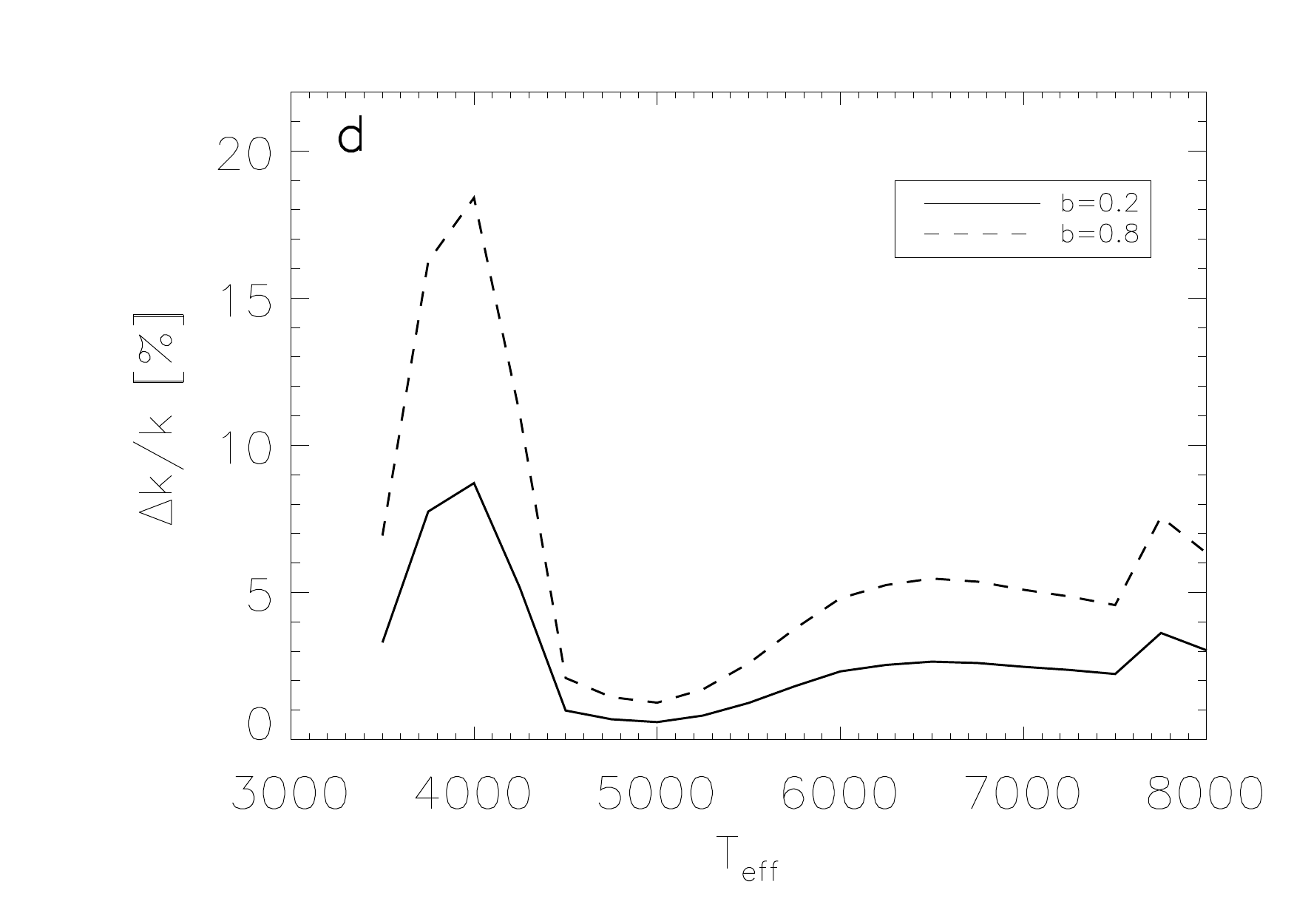}
    \includegraphics[scale=0.33]{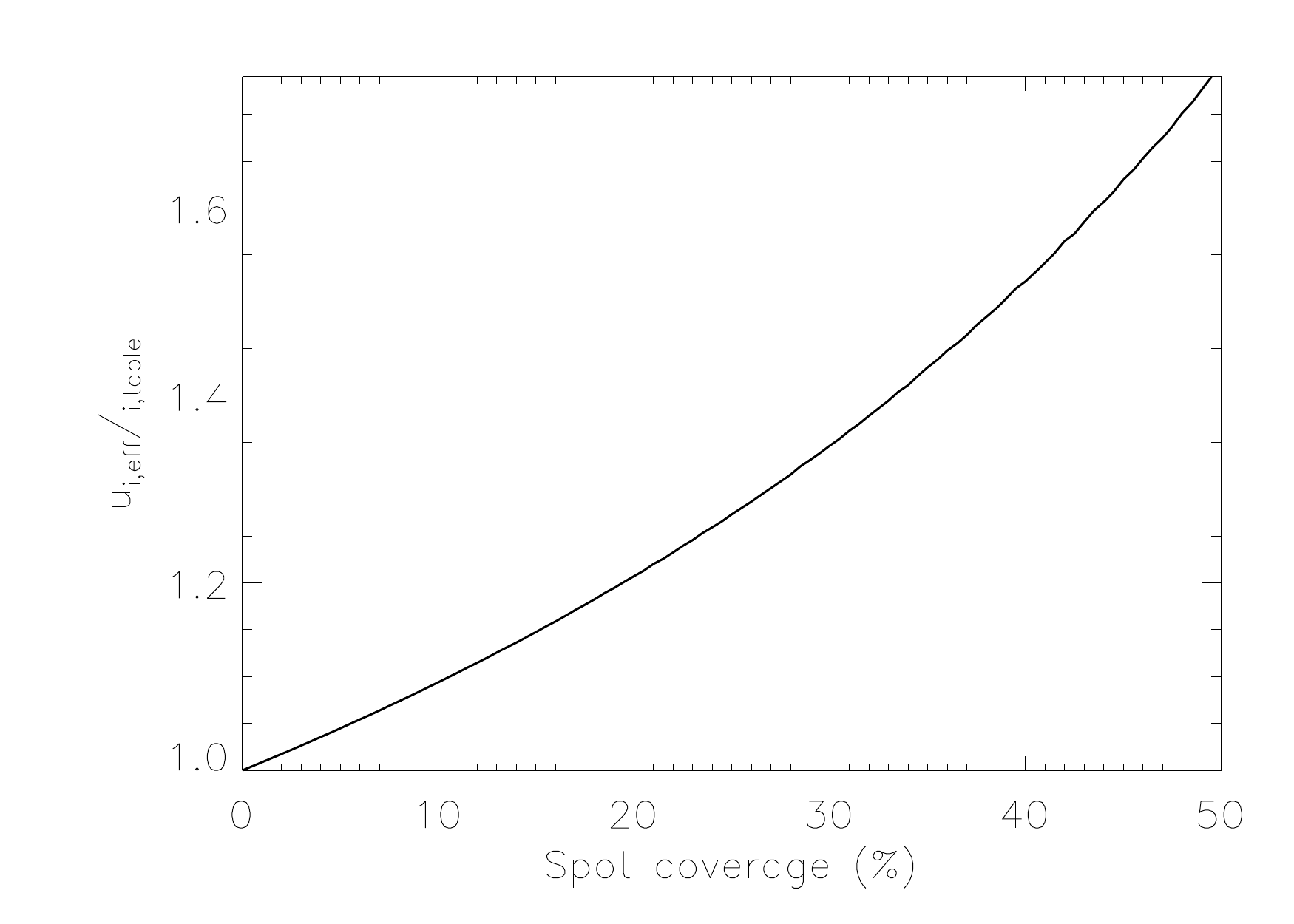}
    \caption{\textit{Left:} Variation of $k \equiv r_\mathrm{p}/R_\star$ from the overshoot given by \citet{sing2010} and the \citet{claret2011} LD coefficients, for a low (continuous line) and grazing (dashed line) impact parameter $b$. \textit{Right:} variations in the observed LD coefficients for varying starspot filling factors for a star with $T_\mathrm{eff} = 5775$~K and $T_\mathrm{spot} = 3775$~K. The positions of the spots were randomly chosen on the visible hemisphere. Figures from \citet{csizmadia2013}, A\&A 549, A9, pp. 4 and 8 (their Figures 2 and 5), reproduced with permission \textcopyright ESO.}
    \label{csizmadia}
\end{figure}

A fit of the LD coefficients to the transit profile is often more desirable than fixing them to theoretical values. As illustrated in Figure \ref{csizmadia}, left panel, this is due to the unavoidable uncertainties in the stellar parameters, as well as to differences in the stellar models that are used. This is particularly the case when stellar activity features produce time-dependent slight deviations of the stellar parameters, that affect the LD coefficients and therefore the transit depth overshoot. When the LD coefficients are included in the transit fit as free parameters, however, the non-linear law often hampers convergence of the MCMC sampling of the transit parameters, and simpler laws need to be used. Starspots were found to potentially produce a few percent variations in the quadratic LD coefficients, as well as up to $\simeq 10\%$ variations in $r_\mathrm{p}/R_\star$, depending on the stellar effective temperature and transit impact parameter \citep{csizmadia2013}. The maximum effect was found around $T_\mathrm{eff} = 4000$~K, grazing transits, and for large spot filling factors, as shown on the right panel in Figure \ref{csizmadia}. Flux-transport models based on the Sun showed that the effect of faculae can also be non-negligible, and that deriving LD curves from the transit fits with minimum residuals induces wavelength-dependent biases due to both spots and facuale \citep{schrijver2020}.

Because of the limited accuracy of the quadratic LD law, some variations were proposed, as well as efficient calculation methods and proper priors on the coefficients to include in the transit fit \citep[e.g.][]{kipping2013,espinoza2016,morello2017,claret2018,maxted2018}. More accurate three-dimensional stellar models also help us better understand the impact of granulation and the behaviour of LD with fundamental parameters \citep[e.g.][]{magic2015}. Due to variations on individual stars and the time dependence of the activity pattern on the stellar photosphere, however, analysing exoplanet transits still requires a compromise between accuracy and computation efficiency when adopting an LD law to implement in an MCMC.

As previously highlighted, LD is a wavelength-dependent effect: it is stronger in the visible, but its effects are also measured in the IR. It has therefore to be taken into account both in transit photometry and spectrophotometry. Observations with the \textit{Hubble Space Telescope} deserve a particular mention: objects which do not lie in the continuous viewing zone of \textit{HST} cannot generally be observed for a full continuous transit. Sometimes, the transit edges are sacrificed in order to obtain a precise measure of the transit depth. As a consequence, LD coefficients cannot be fitted, but are fixed to tabulated values obtained from stellar models at different wavelengths and in the appropriate filters \citep[e.g.][and references therein]{wakeford2019}. The \textit{James Webb Space Telescope} will instead be placed at the Sun-Earth L2 point, and will not suffer from pointing interruptions. For this observatory, the wavelength dependence of the LD coefficients will be fully appreciated. 

\begin{figure}
    \centering
    \includegraphics[scale=0.6]{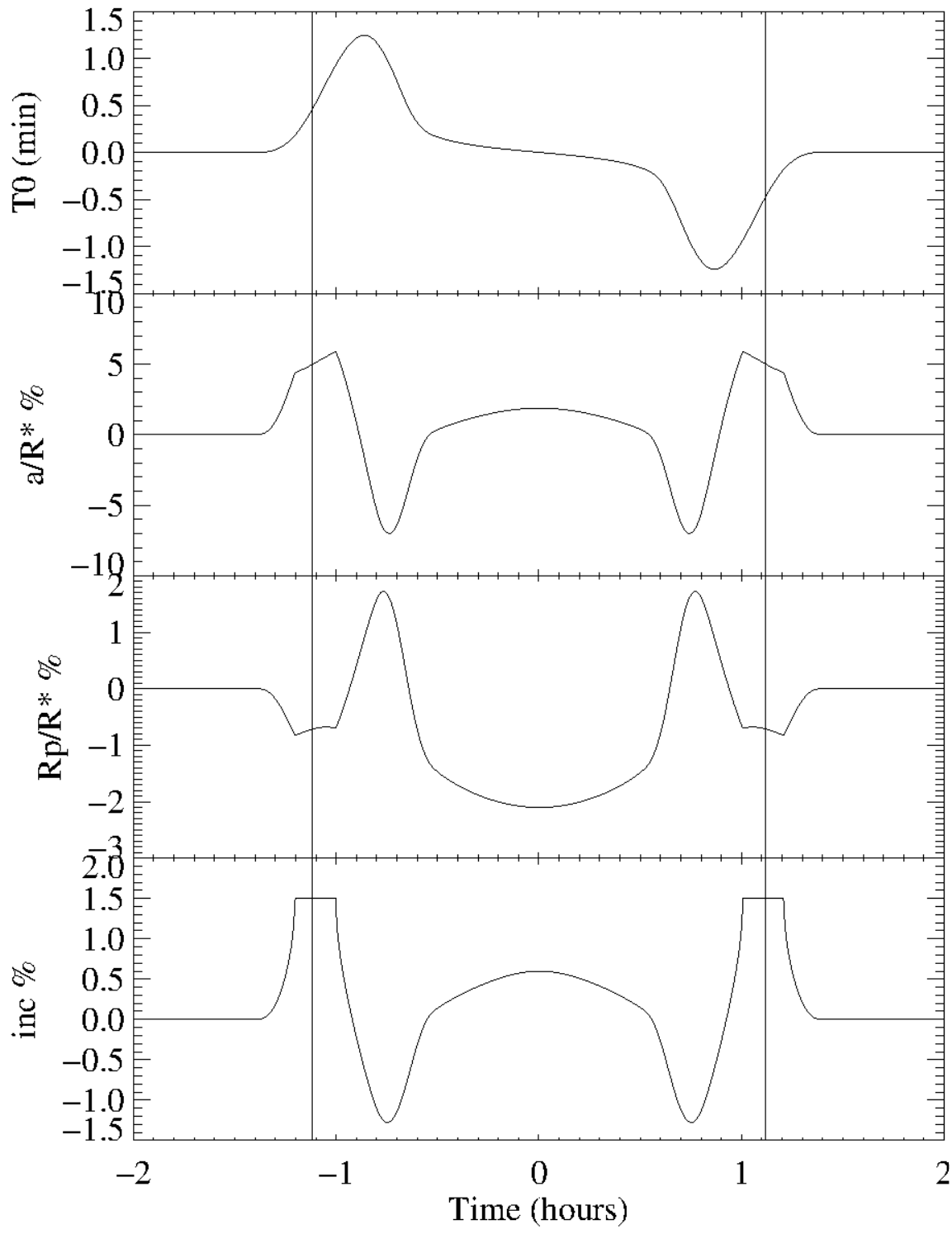}
    \caption{Simulated difference between a transit with and without starspot occultation, as a function of the mid-starspot time compared to mid-transit time, for the parameters of WASP-10b. From top to bottom: transit mid-time, orbital semi-major axis divided by stellar radius, planetary-to-stellar radius ratio and orbit inclination. From \citet{barros2013}, MNRAS 430, 3032–3047 (their Figure 10).}
    \label{barros}
\end{figure}

Stellar activity features have been found to influence the measure of other transit parameters, too. A notable case is the one stellar density, which can be constrained to a high level of precision from planetary transits \citep{pont-moutou2007,sozzetti2007}. Studying the light curve of CoRoT-7b, where the individual transits are not fully resolved, \citet{leger2009} discussed an underestimation of the stellar density from the transit fit compared to the derived spectroscopic value, likely due to stellar variability. Other studies, such as \citet{alonso2009}, \citet{barros2013} and \citet{oshagh2013}, found effects on the apparent transit timing. Figure \ref{barros} shows the simulated difference between a transit with and without starspot occultation, as a function of mid-spot time compared to mid-transit time, for the hot Jupiter WASP-10b \citep{barros2013}. The largest effect is produced when the starspot bump is close to the transit edges, marked with vertical lines. In particular, variations in the mid-transit time (top panel) across multiple transits may mimic transit timing variations (TTVs), which might be misinterpreted as due to non-transiting companions \citep{agol2005}. \citet{barros2013} found hints of starspot contamination by observing a large number of transits with significant  variations in the depth of individual transits.

This latter is another indication that the study of planets around active host stars deserves a special attention. It requires a careful monitoring of the star to evaluate and correct for the contribution of the starspots to the stellar flux due to stellar rotation and their changing characteristics and location. 

\section{Planetary transits and stellar granulation}\label{granulation}

Stellar brightness inhomogeneities caused by stellar convection are another source of noise for transit parameter determination. Granulation produces variations on time scales from tens of minutes to hours. Hot plasma is uplifted towards the stellar surface and produces brighter areas, called granules by \citet{dawes1864} for the first time. Cooler plasma sinks between granules, and produces separating lanes between granules \citep{nordlund1990,nordlund2009}. The first observations on granulation were carried out on the Sun \citep{herschel1801}, but this phenomenon is now thought to be common among Sun-like stars \citep[e.g.][and refereces therein]{chandrasekhar1960,schwarzschild1975,nordlund1982,dravins1990,gilliland2011}.

\subsection{Effect on measured stellar parameters}
The first direct effect of stellar granulation is on the stellar parameters, which are fundamental in order to extract the planetary radius from transit observables. The power of the granulation signal is correlated with the stellar surface gravity $g$: as a star evolves from dwarf toward the giant phase, its $g$ decreases; at the same time, its convection zones deepen and the granulation timescale increases. By measuring the photometric variations on a time scale of 8 hours, \citet{bastien2013} highlighted brightness variations driven by granulation.   The stars of this study span an effective temperature between 4500 to 6750 K, $\log g$ between 2.5 and 4.5 (cgs units), and relative brightness variations lower than $10^{-3}$. Taking advantage of $g$ measurements obtained from \textit{Kepler} asteroseismology observations \citep{chaplin2011}, they derived the ``flicker sequence'' of stellar evolution. This sequence shows a tight correlation between the granulation-induced brightness variation and the stellar $\log g$ (see Figure 1 of \cite{bastien2013}). Within this phase space, \citet{bastien2013} fitted an empirical relation (extended by \cite{bastien2016}) to infer $g$ to within 25\% from the granulation flicker of inactive Sun-like stars, from their main-sequence to their giant phase of evolution.

In addition to surface gravity for the determination of stellar radii, stellar density is also particularly useful to correctly derive transit parameters. This could be carried out  via the so-called ``asterodensity profiling'' \citep{kipping2014AP,sliski2014}. By using 588 catalogued \textit{Kepler} target stars with asteroseismology measurements \citep{huber2013-astero,chaplin2014}, \citet{kipping2014} reported a linear trend between the stellar density and the 8 hr stellar flicker for stars with $4500 \, \mathrm{K} < T_\mathrm{eff} < 6500$~K, $3.25< \log g < 4.43$, and $K_{\rm p}$ magnitude $\leq 14$. The fitted relation has a model error in the stellar density of about 32\%: this is an $\sim 8 \times$ lower precision than asteroseismology, but the relationship can be applied to a $\sim40\times$ wider sample of targets with no asteroseismic measurements. Capitalising on this, \citet{bastien2014} re-estimated the stellar radii of 289 bright ($K_{\rm p}<13$) candidate planet host stars with $4500 \, {\rm K} < T_{\rm eff} < 6650$~K and assessed the impact on the parameters of their planets. They found the $\log g$ derived from broadband photometry and spectroscopy to be systematically lower that the one obtained from the flicker-based calibration ($\sim 0.2$ dex median difference, with RMS $=0.3$ for spectroscopy and 0.4 for broadband photometry). Hence, nearly 50 of the brightest stars in their sample resulted to be subgiants and their radii, together with those of their planets, on average 20\%-30\% larger than previously measured.
 
\subsection{Effect on transit parameters}
Stellar granulation affects the measured transit depth, duration determination of transit ingress and egress, and LD parameters. To quantify the effects, \citet{chiavassa2017} used three-dimensional radiative-hydrodynamical stellar models from the {\sc stagger} grid \citep{magic2013} to model the transit of a hot Jupiter, a hot Neptune and a terrestrial planet in front of the Sun and of a K star.

\begin{figure}
    \centering
    \includegraphics[scale=0.5]{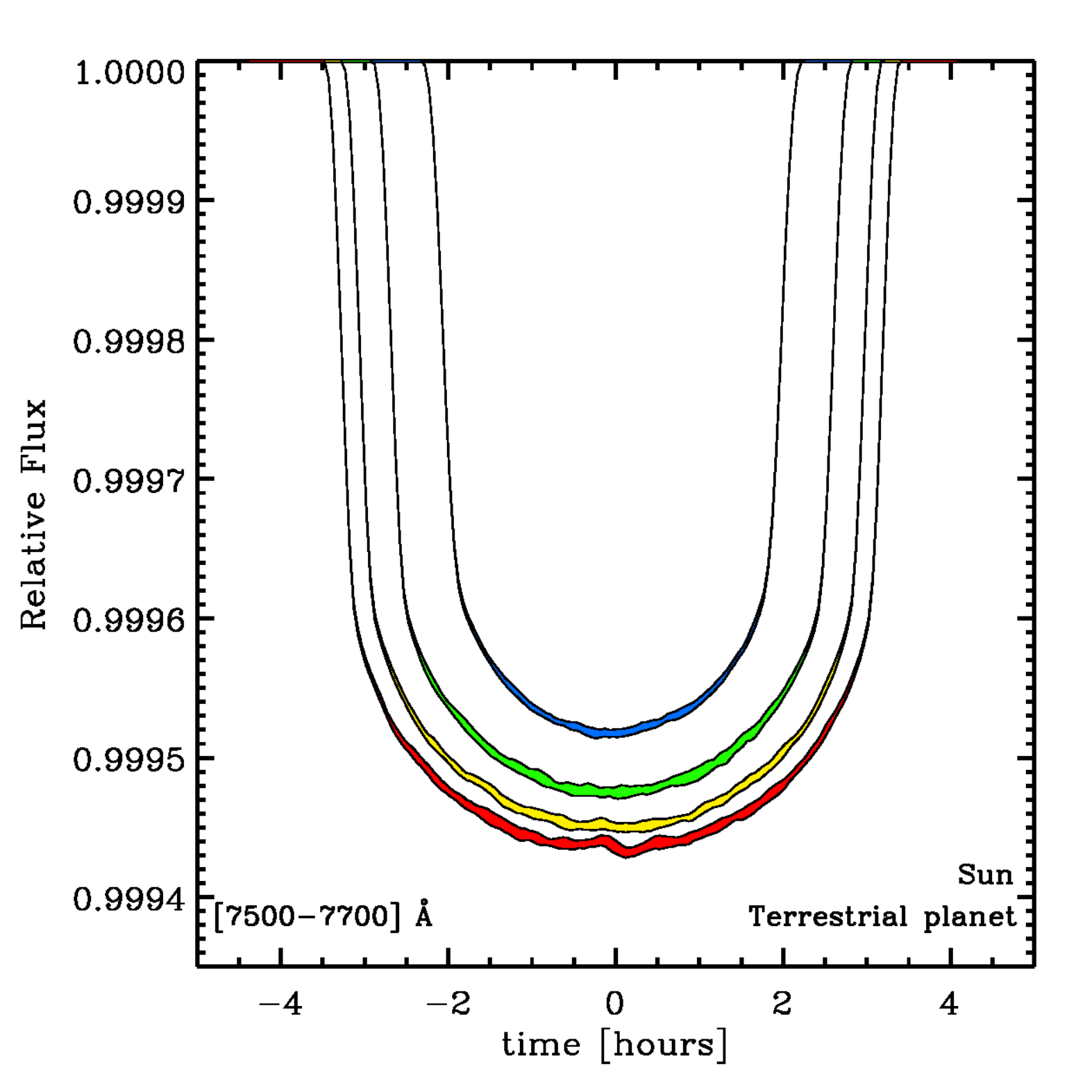}
    \caption{Simulated transit profile variations of a terrestrial planet transiting the Sun, due to different realisations of the modelled granulation pattern. Filled contours indicate the deepest and the dimmest transit profile for different orbital inclinations $i$ in the 750-770 nm~band. Blue corresponds to $i=90.85^\circ$, green to $i=90.65^\circ$, yellow to $i=90.45^\circ$, and red to $i=90.25^\circ$. Credit: \citet{chiavassa2017} (their Figure 11), A\&A 597, A94, p. 10, reproduced with permission \textcopyright ESO.}
    \label{chiavassa}
\end{figure}

The simulations were run by considering several convective turnovers and by taking into account temporal variations in the granulation intensity (e.g. about 10 minutes for the Sun, \cite{nesis2002}). They were also performed in several wavelength windows, covering the bandpasses of most instruments used to observe spectrophotometric transits. The modelled root mean square of the stellar flux in the visible (between 1 and 16 ppm) was close to the observed photometric variability of the SOHO quiet Sun data (10 to 50 ppm: \cite{jenkins2002,frohlich2009}) and was the largest in the visible. Then, different realisations of the stellar irradiance calculated from the granulation patterns were averaged in order to derive the photometric noise they produced. This resulted in detectable variations of $r_{\rm p}/R_\star$, larger for the Sun than for the K star. In the visible, for example, they observed up to $0.90\%$ and up to $0.45\%$ uncertainties in the planet radius for the Sun and the K star simulations, respectively. Also, larger uncertainties were found for terrestrial than for giant planets. Figure \ref{chiavassa} illustrates the case of a terrestrial planet transiting in front of the Sun, in visible wavelengths, for different orbital inclinations. Given the required precision on planetary radii of a few percent at most, in order to be able to significantly constrain planetary core sizes \citep{wagner2011}, such estimates imply that granulation must be considered a non-negligible source of uncertainty. \citet{sulis2020} quantified the effect of granulation on the transit parameters, which  can induce errors of up to 10\% on the ratio between the planetary and stellar radius ($r_\mathrm{p}/R_\star$) for an Earth-sized planet orbiting a Sun-like star. Stellar flicker is already included in the error budget for the \textit{CHEOPS} mission \citep{broeg2013}, and it will be even more so for the \textit{PLATO} mission \citep{rauer2014}, which is designed to achieve a 2\% precision on planetary radii. 

The apparent radius variation was also shown to be dependent on the orbital inclination of the planet, with additional effects on the duration of transit ingress and egress and correlations with the LD parameters. \citet{morris2020} explored the correlations among these effects and found uncertainties on the planetary radius up to 3.6\% in the \textit{PLATO} band, due to granulation and stellar oscillations only. These authors highlighted the importance of independent constraints on the transit impact parameter, or of follow-up observations at longer wavelengths, where LD is weaker, to reduce the significance of this effect. On the other hand, the method used to retrieve the transit parameters seems crucial in the determination of the error budget: Gaussian processes, which are particularly suitable for the modelling of a stochastic phenomenon such as stellar granulation, showed to provide encouraging results with specific kernel choices \citep{barros2020}. 

In the context of the forthcoming \textit{James Webb Space Telescope} \citep[e.g.][]{gardner2006,beichman2014} and the future \textit{Ariel} mission \citep{tinetti2018}, the potential impact of stellar granulation and oscillations on transit spectrophotometry has started to be evaluated. \citet{sarkar2018} reported that the expected effect is weak, especially in the near-IR. There, its impact could even be negligible, but with a dependency on the atmospheric scale height and the transit duration. As a case which is most likely to be impacted by granulation, these authors mention the one of `a terrestrial planet with a secondary atmosphere orbiting a nearby Sun-like star on a long period' \citep[][Sarkar et al. 2018, p. 2877]{sarkar2018}.

\section{Conclusions}\label{concl}
In this review, we presented the main issues caused by stellar activity on exoplanet transit observations, as well as the main approaches which are used to correct for them. In several cases, the combination of ultra-precise measurements and theoretical modelling is required to precisely quantify the possible biases on planetary parameters. Depending on the observations and the scientific goal, complementary observations can provide a useful way to better assess the effects of stellar activity. On the other hand, attention to the determination of the physical properties of stellar activity features such as starspots, faculae and granulation and of their temporal evolution has grown in the past few years. The goal of a few percent precision in planet radii, and few tens of parts per million in transmission spectroscopy, is challenged by the imprints of stellar activity on the observables, and requires these phenomena to be taken into account as an unavoidable component of star-planet observations.

As we push our search towards terrestrial planets, the understanding of stellar activity and its proper correction becomes an increasingly necessary complement to improved instrument performance. The best outcomes will then result from synergies between different observation techniques and modelling perspectives. Thanks to these, we will be able to obtain as much precise as possible insights into the physical properties of the most interesting discovered systems. At the same time, this race towards small size planets with precise and accurate parameters will help refine our knowledge and understanding of stars.

\acknowledgements
GB acknowledges support from CHEOPS ASI-INAF agreement n. 2019-29-HH.0. MD acknowledges support by CNES, focused on the PLATO mission. 

%%-----------------------------
%%      your bibliography
%%-----------------------------
\printbibliography
\end{document}